\documentclass{aa} 
\usepackage{graphicx}
\usepackage{dblfloatfix}
\usepackage{txfonts}
\usepackage{hyperref}
\begin{document}

   \title{Carbon depletion observed inside T Tauri inner rims:}

   \subtitle{Formation of icy, kilometer size planetesimals by 1 Myr}

   \author{M. K. McClure 
          \inst{1}
          }

   \institute{$^1$Marie Curie Postdoctoral Fellow,
Anton Pannekoek Institute for Astronomy,
Universiteit van Amsterdam,
Science Park 904, 
1098 XH Amsterdam, Netherlands
              \email{m.k.mcclure@uva.nl}
             }

   \date{Received -; accepted July 2019}

 
  \abstract
   {The carbon content of protoplanetary disks is an important parameter to characterize planets formed at different disk radii. There is some evidence from far-infrared and submillimeter observations that gas in the outer disk is depleted in carbon, with a corresponding enhancement of carbon-rich ices at the disk midplane. Observations of the carbon content inside of the inner sublimation rim could confirm how much carbon remains locked in kilometer size bodies in the disk.}
   {I aim to determine the density, temperature, and carbon abundance inside the disk dust sublimation rim in a set of T Tauri stars with full protoplanetary disks.}
   {Using medium-resolution, near-infrared (0.8 to 2.5 $\mu$m) spectra and the new GAIA DR 2 distances, I self-consistently determine the stellar, extinction, veiling, and accretion properties of the 26 stars in my sample. From these values, and non-accreting T Tauri spectral templates, I extract the inner disk excess of the target stars from their observed spectra . Then I identify a series of C$^0$ recombination lines in 18 of these disks and use the CHIANTI atomic line database with an optically thin slab model to constrain the average $n_e$, $T_e$, and $n_C$ for these lines in the five disks with a complete set of lines. By comparing these values with other slab models of the inner disk using the Cloudy photoionization code, I also constrain $n_H$ and the carbon abundance, $X_C$, and hence the amount of carbon `missing' from the slab. For one disk, DR Tau, I use relative abundances for the accretion stream from the literature to also determine $X_{Si}$ and $X_N$.}
   {The inner disks modeled here are extremely dense ($n_H\sim10^{16}$ cm$^{-3}$), warm ($T_e\sim4500$ K), and moderately ionized (log$X_e\sim3.3$). Three of the five modeled disks show robust carbon depletion up to a factor of 42 relative to the solar value. I discuss multiple ways in which the `missing' carbon could be locked out of the accreting gas. Given the high-density inner disk gas, evidence for radial drift, and lack of obvious gaps in these three systems, their carbon depletion is most consistent with the `missing' carbon being sequestered in kilometer size bodies. For DR Tau, nitrogen and silicon are also depleted by factors of 45 and 4, respectively, suggesting that the kilometer size bodies into which the grains are locked were formed beyond the N$_2$ snowline. I explore briefly what improvements in the models and observations are needed to better address this topic in the future.}
  { - }


   \maketitle
%

\section{Introduction}
The distribution of carbon in protoplanetary disks is critical to understanding the organic content of exoplanets and therefore the origins of life. The bulk C/O ratio of planetary atmospheres may be linked back to the solid material accreted in the protoplanetary disk \citep{madhusudhan12b}. Recent observations of protoplanetary disks demonstrate that CO may be more depleted from the gas phase of the outer tens of AU than can be explained simply by photodissociation and freeze-out. In the three disks with gas masses derived using both CO and HD, the gas masses derived from CO are lower by factors of 5 - 100 \citep{bergin+13, mcclure+16}. Full chemical models of spatially resolved ALMA CO isotopologue observations in TW Hya confirm that CO is chemically depleted by a factor of 100 in the outer disk, even taking into account photodissociation and freeze-out \citep{schwarz+16a}. Additional chemical modeling of single-dish observations of the atomic [C$^0$] 492.161 GHz line, which originates in the photodissociated layer of the disk atmosphere, demonstrates that the depletion seen in CO can be generalized to the bulk carbon budget as well \citep{kama+16b}. These authors find that the outer disk gas phase CO and carbon depletion can be explained by freeze-out of volatile carbon, in the form of CO ice, with subsequent grain surface reactions converting these ices into less volatile ices, like CO$_2$ or methanol, which keeps the gas phase CO abundance low interior to the CO snowline \citep{schwarz+16a,vanthoff2017}. 

Transformation of solid CO into a more refractory form of carbon on dust grains could enhance the amount of carbon-rich solids available to make planetary cores. If the coagulation of such grains into large bodies beyond the innermost carbon-bearing solid sublimation radius is efficient, then the carbon would be `locked' into the disk and the gas phase carbon abundance would remain low. In contrast, inefficient solid growth would lead to smaller bodies that radially drift towards the star, returning the `missing' carbon to the gas phase in the inner disk when the last icy grains have sublimated. Further modeling of HD, spatially resolved CO isotopologues, and resolved continuum observations in TW Hya suggests that up to 2.4 M$_{\oplus}$ of dust may have grown to larger than centimeter sizes just outside of 5 AU \citep{zhang2017}, which could permanently sequester volatile carbon. However, that analysis depends on the dust properties assumed for the continuum, and observations at sufficiently high spatial resolution are difficult to obtain at the distances of typical star forming regions, >100 pc. To measure the degree of volatile locking in a larger sample, a better solution is to measure the gas phase carbon abundances inside of the silicate sublimation radius, where the gas is thought to be optically thin for mass accretions rates of less than 10$^{-7}$ M$_{\odot}$ yr$^{-1}$, based on models using molecular gas opacities \citep{muzerolle04}. For high-mass Herbig AeBe stars, these abundances can be inferred from stellar photosphere abundances because the photospheric convective layer is negligible, so recently accreted gas remains at the surface \citep{kama+16b}. However, the deeper convective layer in low-mass T Tauri stellar atmospheres prevents their stellar abundances to be used in this way to infer inner disk abundances. Disk abundance measurements of T Tauri stars must therefore be made through observations of the accretion columns \citep{ardila2013} or the inner disk gas directly.

In 2013, I identified neutral atomic carbon emission in a set of accreting T Tauri stars, as part of a pilot study of the inner dust rim \citep{mcclure+13a}. These lines appeared in the strongest accretor in our sample, DR Tau, and grew weaker for the lower mass accretion rate objects. To investigate these lines and the inner dust rim, I observed an enlarged sample of T Tauri stars in Taurus from 0.8 to 2.5 $\mu$m. Here I present an initial analysis of the carbon line fluxes in this sample, focusing on the temperatures and densities required to excite them. The sample, extraction of the inner disk excess, and line fluxes are described in \S \ref{obs}, the analysis of the temperatures, densities, carbon ionization fraction, and carbon abundance in \S \ref{mod_nete}, and the implications for the inner disk chemistry and formation of kilometer size bodies are discussed in \S \ref{discussion}.


\section{Observations and data analysis}
\label{obs}
The full sample contains 26 well-studied accreting, `classical' T Tauri stars (CTTS) in the nearby Taurus-Auriga molecular cloud complex \citep[$\sim$140 pc,][]{kenyon+94}. The initial selection criteria were that the stars must be 1) single within a detection limit of $\delta$mK = 2 at separations greater than 20 mas, with 2) no evidence for gaps or radial structure based on their infrared SEDS, and 3) have {\it Spitzer} Infrared Spectrograph (IRS) spectra. Subsequent criteria were that the disks needed to cover a broad range of spectral types and mass accretion rates, based on previous optical publications, e.g. \citet{kh95}. Based on these optical spectral types, I also observed bright non-accreting, `weak-line' T Tauri stars (WTTS) for each target spectral type to act as photospheric templates for the CTTS.  Data for ten stars were already published in \citet{mcclure+13a} as part of a pilot study.

\subsection{Observations}
The spectra were obtained with SpeX \citep{rayner+03} at the NASA Infrared Telescope Facility (IRTF) on December 1-3, 2010 and January 8-12, 2013. I observed the targets with the short wavelength, cross-dispersed mode (SXD) with the 0\farcs3 by 15\farcs0 slit (R=2000) from 0.8 to 2.5 $\mu$m with integration times that were selected to produce a SNR > 100 at H band. The data were obtained with the slit rotated to the parallactic angle and in an ABBA nod pattern. Full details of the data reduction are given in \citet{mcclure+13a}. In brief, the spectra were reduced with the standard Spextool package \citep{cushing+04,vacca+03}, resulting in sky subtracted, telluric corrected, and flux calibrated spectra. 

\subsection{Data analysis: Inner disk excess extraction}
\label{stellparm}
The spectra of the WTTS and CTTS were corrected for radial velocities by shifting them to match the position of the strong 1.31 $\mu$m Al I doublet seen in the velocity-corrected IRTF spectral library G, K, and M dwarfs \citep{rayner+09}. Then the WTTS were extinction-corrected to match the continuum between the molecular bands shortward of 1.1 $\mu$m of the dwarf spectrum of the same spectral type from the IRTF standards library. Using the analysis routines from \citet{mcclure+13a} and the additional WTTS spectra, I constructed equivalent width, $W_{\lambda}$, versus SpT trends for the WTTS in this sample and compared them with the IRTF library dwarf and giant equivalent widths.  I then computed ratios between equivalent widths of nearby lines to use in determining the veiling-independent spectral type. However, some of the line ratios used in \citet{mcclure+13a} showed a local maximum around K0; given the larger range of CTTS and WTTS spectral types (F8-M4), it was necessary to select additional ratios.

Having determined the spectral types of the CTTS, I used the $W_{\lambda}$ measurements for individual lines in the CTTS and WTTS templates to determine the amount of veiling in these lines. The veiling measurements and continuum WTTS fluxes were used to compute the extinction for each CTTS as in \citet{mcclure+13a}. After extinction correction, the WTTS photospheric template is offset from the CTTS spectrum using the average measured veiling between 1.0 and 1.35$\mu$m. The stellar luminosity, $L_{*}$, is calculated from the correctly offset WTTS template, using the colors and bolometric corrections given in \citet{kh95} and new stellar distances from GAIA DR2 \citep{bailer-jones18} where available or the standard value of 140 pc otherwise \citep{kenyon+94}. Together with the effective temperatures given by the near-IR spectral types, these luminosities imply a stellar radius $R_{*}$. These parameters are compared with the Siess evolutionary tracks \citep{siess+00} to derive the stellar mass, $M_{*}$. From the excess emission spectrum, I measure the luminosity in the H$^0$ Br$\gamma$ line and use the trend from \citet{muzerolle03a} to compute the accretion luminosity, $L_{acc}$, and mass accretion rate, $\dot M$. The filling factor of the accretion shock on the stellar surface can be computed from $L_{acc}$, $L_{*}$, and $T_{eff}$ assuming an average temperature of 8000 K in the post-shock region at the stellar surface: $f = \frac{L_{acc}/L_{*}}{L_{acc}/L_{*} + (8000/T_{eff})^4}$. The major caveat on this method for calculation of the accretion column filling factor is that the Br$\gamma$ emission potentially originates in three regions: the accretion shock at the stellar surface, the accretion columns, and the inner disk. A more accurate estimate of the filling factors can be obtained by fitting accretion shock models to the UV excess \citep{ingleby+13}; however I cannot make this calculation from the present data.  The derived stellar and accretion parameters are given in Table \ref{table:1}. 

Subtraction of the scaled WTTS photospheric template from the observed extinction-corrected CTTS spectrum reveals the combined excess emission from the accretion shock, inner gaseous disk, and dust sublimation rim. 

\begin{table*}
\scriptsize
\caption{Targets and derived stellar parameters}             
\label{table:1}      
\centering                          
\begin{tabular}{c c c c c c c c c c c c c c c}        
\hline\hline                 
 \\
Star & Date & SpT & T$_{eff}$ & A$_V$   & d & L$_{*}$ & L$_{Br\gamma}$ & L$_{acc}$ & $f_{sh}$ & R$_{*}$ & M$_{*}$ & age & $\dot M$ & $\overline{r_J}$ \\
     &      & (IR)&  [K]      & [mag] & [pc] & [L$_{\odot}$]& [L$_{\odot}$]    & [L$_{\odot}$]&           & [R$_{\odot}$]  & [M$_{\odot}$] &  [Myr] & [M$_{\odot}$ yr$^{-1}$] &  \\
\hline
\\
BP Tau & 2010-12-02	&  M0V & 3850	&  0.5 $\pm$ 1.4 & 128.6 $\pm$ 1.0	& 0.84	& 3.2$\times$10$^{-5}$	& 6.0$\times$10$^{-2}$	& 3.8$\times$10$^{-3}$	& 2.06	& 0.562	& 1.3	& 8.8$\times$10$^{-9}$	&  0.2  \\
CI Tau & 2010-12-02	&  K7V & 4060	&  0.5 $\pm$ 1.0 & 158.0 $\pm$ 1.2	& 0.73	& 2.0$\times$10$^{-4}$	& 5.6$\times$10$^{-1}$	& 4.9$\times$10$^{-2}$	& 1.72	& 0.765	& 2.7	& 5.0$\times$10$^{-8}$	&  0.4  \\
CW Tau & 2013-01-13	&  K0V & 5250	&  6.4 $\pm$ 0.4 & 131.9 $\pm$ 0.7	& 0.86	& 2.3$\times$10$^{-4}$	& 7.3$\times$10$^{-1}$	& 1.4$\times$10$^{-1}$	& 1.13	& 1.049	& 27.8	& 3.1$\times$10$^{-8}$	&  3.2  \\
CX Tau & 2013-01-09	&  M2V & 3580	&  0.4 $\pm$ 1.0 & 127.5$^{+0.7}_{-0.6}$	& 0.40	& 1.2$\times$10$^{-6}$	& 9.9$\times$10$^{-4}$	& 9.9$\times$10$^{-5}$	& 1.64	& 0.393	& 1.8	& 1.7$\times$10$^{-10}$	&  0.1  \\
DE Tau & 2010-12-03	&  M2V & 3580	&  1.1 $\pm$ 0.8 & 126.9 $\pm$ 1.1	& 0.55	& 3.9$\times$10$^{-5}$	& 7.8$\times$10$^{-2}$	& 5.7$\times$10$^{-3}$	& 1.94	& 0.395	& 1.3	& 1.5$\times$10$^{-8}$	&  0.4  \\
DK Tau & 2013-01-11	&  K7V & 4060	&  0.9 $\pm$ 0.7 & 128.1 $\pm$ 1.0	& 0.93	& 2.6$\times$10$^{-5}$	& 4.6$\times$10$^{-2}$	& 3.3$\times$10$^{-3}$	& 1.95	& 0.753	& 2.0	& 4.8$\times$10$^{-9}$	&  0.3  \\
DL Tau & 2013-01-12	&  K7V & 4060	&  1.6 $\pm$ 0.5 & 158.6 $\pm$ 1.2	& 0.37	& 3.1$\times$10$^{-4}$	& 9.4$\times$10$^{-1}$	& 1.4$\times$10$^{-1}$	& 1.24	& 0.804	& 7.8	& 5.8$\times$10$^{-8}$	&  2.1  \\
DO Tau & 2013-01-11	&  M0V & 3850	&  3.6 $\pm$ 0.5 & 138.8 $\pm$ 1.0	& 0.58	& 1.9$\times$10$^{-4}$	& 5.4$\times$10$^{-1}$	& 4.8$\times$10$^{-2}$	& 1.72	& 0.569	& 2.0	& 6.6$\times$10$^{-8}$	&  2.0  \\
DR Tau & 2010-12-02	&  M0V & 3850	&  2.1 $\pm$ 0.5 & 194.6$^{+2.5}_{-2.4}$	& 1.12	& 8.0$\times$10$^{-4}$	& 2.8 	& 1.2$\times$10$^{-1}$	& 2.39	& 0.558	& 0.9	& 4.8$\times$10$^{-7}$	&  2.7  \\
DS Tau & 2010-12-04	&  M0V & 3850	&  1.2 $\pm$ 1.1 & 158.4 $\pm$ 1.1	& 1.09	& 5.2$\times$10$^{-5}$	& 1.0$\times$10$^{-1}$	& 5.0$\times$10$^{-3}$	& 2.34	& 0.559	& 1.0	& 1.7$\times$10$^{-8}$	&  0.1  \\
FQ Tau & 2013-01-13	&  M4V & 3370	&  1.4 $\pm$ 1.1 & 140  & 0.28	& 1.5$\times$10$^{-6}$	& 1.3$\times$10$^{-3}$	& 1.4$\times$10$^{-4}$	& 1.57	& 0.3	& 1.9	& 2.6$\times$10$^{-10}$	&  0.1  \\
FS Tau & 2013-01-09	&  M0V & 3850	&  5.5 $\pm$ 1.1 & 140 & 1.03	& 2.9$\times$10$^{-5}$	& 5.2$\times$10$^{-2}$	& 2.7$\times$10$^{-3}$	& 2.28	& 0.561	& 1.0	& 8.5$\times$10$^{-9}$	&  0.3  \\
FT Tau & 2013-01-09	&  M4V & 3370	&  1.4 $\pm$ 0.8 & 127.3$^{+0.9}_{-0.8}$	& 0.25	& 4.0$\times$10$^{-5}$	& 8.0$\times$10$^{-2}$	& 1.0$\times$10$^{-2}$	& 1.46	& 0.298	& 2.2	& 1.6$\times$10$^{-8}$	&  0.3  \\
FX Tau & 2013-01-11	&  M0V & 3850	&  2.7 $\pm$ 1.4 & 140 & 1.31	& 1.8$\times$10$^{-5}$	& 2.8$\times$10$^{-2}$	& 1.1$\times$10$^{-3}$	& 2.57	& 0.555	& 0.8	& 5.1$\times$10$^{-9}$	&  0.1  \\
FZ Tau & 2013-01-10	&  M0V & 3850	&  6.5 $\pm$ 0.7 & 129.6 $\pm$ 1.3	& 0.93	& 5.6$\times$10$^{-4}$	& 2.2 	& 1.1$\times$10$^{-1}$	& 2.18	& 0.561	& 1.1	& 3.5$\times$10$^{-7}$	&  1.1  \\
GI Tau & 2013-01-11	&  M0V & 3850	&  3.7 $\pm$ 1.9 & 130.0 $\pm$ 0.8	& 0.60	& 2.7$\times$10$^{-5}$	& 4.9$\times$10$^{-2}$	& 4.4$\times$10$^{-3}$	& 1.73	& 0.568	& 1.9	& 5.9$\times$10$^{-9}$	&  0.4  \\
GK Tau & 2013-01-11	&  K7V & 4060	&  2.0 $\pm$ 0.6 & 128.8 $\pm$ 0.7	& 0.91	& 3.7$\times$10$^{-5}$	& 7.3$\times$10$^{-2}$	& 5.3$\times$10$^{-3}$	& 1.92	& 0.753	& 2.0	& 7.4$\times$10$^{-9}$	&  0.5  \\
GN Tau & 2013-01-13	&  M2V & 3580	&  4.5 $\pm$ 0.7 & 140 							& 0.93	& 8.1$\times$10$^{-5}$	& 1.9$\times$10$^{-1}$	& 8.0$\times$10$^{-3}$	& 2.51	& 0.395	& 0.9	& 4.7$\times$10$^{-8}$	&  0.4  \\
Haro 6-28 & 2013-01-13	&  K5V & 4350	&  3.3 $\pm$ 0.6 & 140 & 0.27	& 4.7$\times$10$^{-6}$	& 5.3$\times$10$^{-3}$	& 1.7$\times$10$^{-3}$	& 0.91	& 0.831	& 25.4	& 2.3$\times$10$^{-10}$	&  0.3  \\
HP Tau & 2013-01-13	&  K0V & 5250	&  3.0 $\pm$ 0.5 & 176.4$^{+3.4}_{-3.3}$	& 4.70	& ...	& ...	& ...	& 2.62	& 1.936	& 4.7	& ...	&  0.8  \\
HP Tau G3	& 2013-01-09	&  M0V & 3850	&  2.7 $\pm$ 1.5 & 156.5 $\pm$ 2.1	& 0.91	& ...	& ...	& ...	& 2.15	& 0.561	& 1.2	& ...	&  0.0  \\
HQ Tau & 2013-01-09	&  K2V & 4900	&  2.8 $\pm$ 0.8 & 158.2$^{+5.6}_{-5.2}$	& 4.01	& 4.8$\times$10$^{-6}$	& 5.1$\times$10$^{-3}$	& 1.8$\times$10$^{-4}$	& 2.78	& 1.977	& 2.7	& 2.8$\times$10$^{-10}$	&  0.1  \\
IQ Tau & 2013-01-13	&  M0V & 3850	&  2.4 $\pm$ 1.3 & 130.8 $\pm$ 1.1	& 1.00	& 2.0$\times$10$^{-5}$	& 3.3$\times$10$^{-2}$	& 1.8$\times$10$^{-3}$	& 2.25	& 0.561	& 1.1	& 5.3$\times$10$^{-9}$	&  0.1  \\
IRAS 04125+2902 & 2013-01-09	&  M0V & 3850	&  2.9 $\pm$ 1.5 & 159.2 $\pm$ 1.7	& 0.57	& 3.6$\times$10$^{-7}$	& 1.9$\times$10$^{-4}$	& 1.8$\times$10$^{-5}$	& 2.74	& 0.569	& 2.1	& 3.7$\times$10$^{-11}$	&  0.0  \\
IRAS 04303+2240 & 2013-01-11	&  M0V & 3850	&  9.0 $\pm$ 1.1 & 147.5$^{+6.5}_{-5.9}$	& 1.92	& 9.0$\times$10$^{-5}$	& 2.1$\times$10$^{-1}$	& 5.8$\times$10$^{-3}$	& 3.11	& 0.552	& 0.6	& 4.7$\times$10$^{-8}$	&  0.4  \\
RY Tau & 2013-01-13	&  G0V & 6030	&  3.2 $\pm$ 0.7 & 443.7$^{+55.1}_{-44.3	}$	& 141.19	& 5.5$\times$10$^{-3}$	& 2.1$\times$10$^1$	& 4.6$\times$10$^{-2}$	& 10.89	& 4.629	& 0.5	& 2.0$\times$10$^{-6}$	&  0.2 \\
\\
\hline                                   
\end{tabular}
 \\
 Notes: Columns are defined as: (1) central star name, (2) date of observation, (3) stellar spectral type, (4) effective stellar temperature, (5) 
 extinction along the line of sight, (6) distance to the star using either the GAIA DR2 supplemental catalog from \citet{bailer-jones18} or the default distance of the Taurus star-forming region of 140 pc \citep{kenyon+94},  (7) stellar luminosity, (8) luminosity in the H$^0$ Br$\gamma$ line, (9) the accretion luminosity derived from Br$\gamma$, 
 (10) the filling factor of the accretion shock on the central star, (11) the stellar radius, (12) the stellar mass taken from the \citet{siess+00} evolutionary tracks, (13) the stellar age from the same tracks, (14) the mass accretion rate derived from Br$\gamma$, (15) the mean J-band veiling. All stellar quantities are derived here using the procedure from \citet{mcclure+13a}. Missing mass accretion rates resulted in two sources, HP Tau and HP Tau G3, for which the Br$\gamma$ line was in absorption even after subtracting the stellar photosphere.
\end{table*}

\subsection{Data analysis: Carbon line identification and fluxes}
\begin{figure*}
   \centering
   \includegraphics[width=17cm]{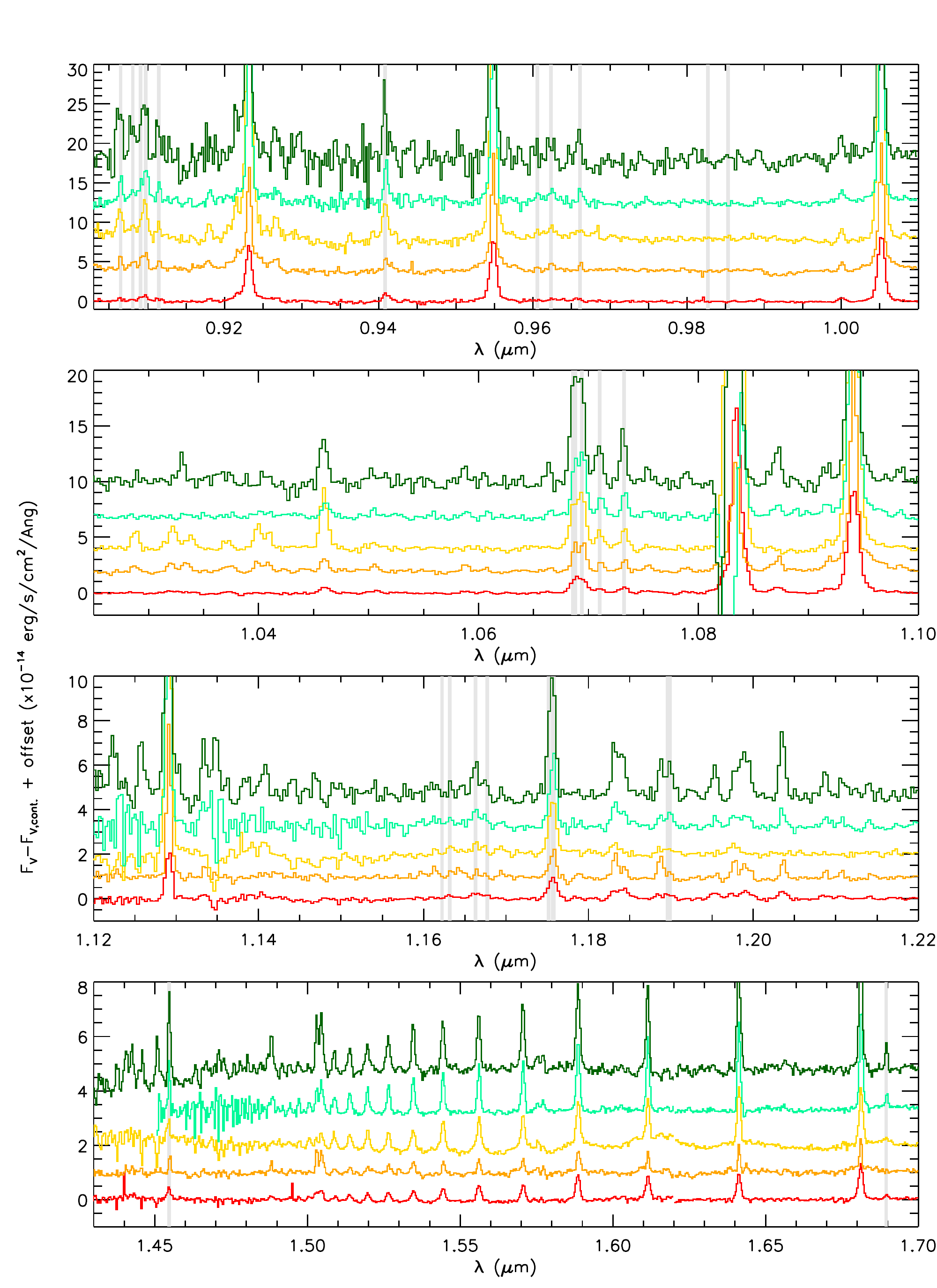}
   \caption{Continuum subtracted excess emission covering the C$^0$ lines
               for the five disks with the best S\/N spectrum: FZ Tau, DR Tau, CW Tau, DO Tau, and DL Tau (from top). Grey vertical lines indicate positions of C$^0$ lines of interest. Note the forbidden lines at 0.982 and 0.985 $\mu$m are not detected, in contrast with the permitted lines.}
              \label{contsubs_strong}%
    \end{figure*}

The excess, which is a mostly smooth continuum overlaid with emission lines, is fit with a polynomial that is subsequently subtracted off. The continuum subtracted excesses for the five disks with the highest SNR spectra are plotted in Figure \ref{contsubs_strong}. In the excess emission, I detect a series of emission lines at 5-20\% of the continuum level. Using the NIST atomic line database \citep{NIST_ASD} and the list of lines in \citet{escalante90} and \citet{haris_2017}, I identify this emission as carbon recombination lines arising in a predominantly neutral carbon reservoir. The atomic data for all detected transitions are listed in Appendix A. The lines are permitted, with upper states energies of 8-9 eV. They are not seen in the WTTS stars used as photospheric templates, nor in similar IRTF SXD, Magellan FIRE, or VLT X-Shooter spectra of transitional disks or pre-transitional disks, which have reduced gas and dust masses interior to some disk radius \citep{vacca_sandell11, espaillat+14}.

\begin{figure}
   \centering
   \includegraphics[width=\hsize]{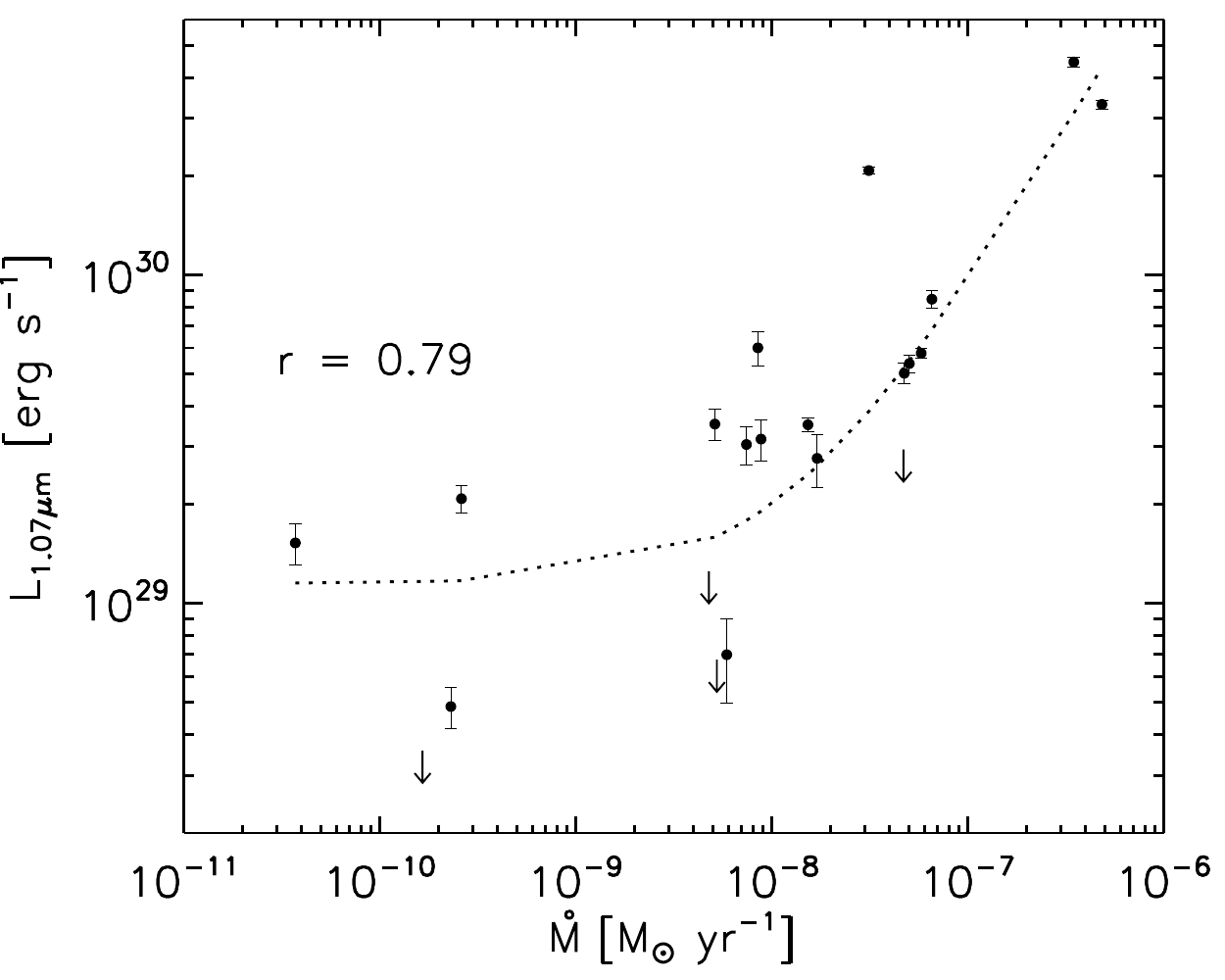}
      \caption{Integrated luminosity of the three strongest lines in the main 1.07 $\mu$m C$^0$ complex plotted against the mass accretion rate, $\dot M$, calculated from the the H$^0$ Br$\gamma$ line at 2.166 $\mu$m. Uncertainties in the C$^0$ line luminosities are included, along with a linear regression fit to the data (dotted line). The Pearson correlation coefficient is 0.79, indicating a high likelihood of association with $\dot M$.}
         \label{cibr}
   \end{figure}

For all of the disks, I calculate an integrated flux for the main line clusters (see Table \ref{table:2}). Since many of the C$^0$ lines are blended, this flux is a manual integration of multiple lines rather than separate gaussian fits. The strongest set of the series is a five-line complex at 1.069 $\mu$m; the integrated flux of this set of lines correlates with the flux of Br$\gamma$, an accretion indicator, suggesting that it is either associated with the accretion columns/shock or with the innermost gaseous disk inside the dust sublimation rim (Fig. \ref{cibr}). This region is predicted to be optically thin for higher mass Herbig AeBe stars with mass accretion rates of less than $\sim$1$\times$10$^{-7}$ $M_{\odot}$ yr$^{-1}$ \citep{muzerolle04}.

\begin{table*}
\scriptsize
\caption{Integrated fluxes of C$^0$ line complexes}             
\label{table:2}      
\centering                          
\begin{tabular}{c c c c c c c c c c c c c c}        
\hline\hline                 
 \\
Star & 0.908 $\mu$m &  & 0.94 $\mu$m &   & 1.069 $\mu$m &  & 1.176 $\mu$m &  & 1.45 $\mu$m &   & 2.166 $\mu$m &   & Notes  \\
        &   [erg/cm$^{2}$/s]   & S/N &  [erg/cm$^{2}$/s]   & S/N  & [erg/cm$^{2}$/s]   & S/N & [erg/cm$^{2}$/s]   & S/N &   [erg/cm$^{2}$/s]   & S/N  & [erg/cm$^{2}$/s]   & S/N  &   \\
\hline
\\
BP Tau & 1.5$\times$10$^{-13}$ & 5.9 & 1.3$\times$10$^{-13}$ &  3$\sigma$ & 1.6$\times$10$^{-13}$ & 7 & 5.9$\times$10$^{-14}$ & 6.4 & 4.1$\times$10$^{-14}$ &  3$\sigma$ & 1.3$\times$10$^{-13}$ & 16.8 &  \\
CI Tau & 1.2$\times$10$^{-13}$ & 6.6 & 8.5$\times$10$^{-14}$ &  3$\sigma$ & 1.8$\times$10$^{-13}$ & 16.5 & 5.7$\times$10$^{-14}$ & 7.8 & - & - & 3.1$\times$10$^{-13}$ & 35.5 &  \\
CW Tau & 1.0$\times$10$^{-12}$ & 16.9 & 3.7$\times$10$^{-13}$ & 8.9 & 1.0$\times$10$^{-12}$ & 44.8 & 2.8$\times$10$^{-13}$ & 34 & 1.4$\times$10$^{-13}$ & 8.7 & 6.3$\times$10$^{-13}$ & 97.8 &  \\
CX Tau & 2.8$\times$10$^{-14}$ & 5.4 & 5.7$\times$10$^{-14}$ &  3$\sigma$ & 2.3$\times$10$^{-14}$ &  3$\sigma$ & 1.3$\times$10$^{-14}$ &  3$\sigma$ & 9.0$\times$10$^{-15}$ &  3$\sigma$ & 5.9$\times$10$^{-14}$ & 10.5 &  \\
DE Tau & 2.0$\times$10$^{-13}$ & 10 & 5.1$\times$10$^{-14}$ &  3$\sigma$ & 1.8$\times$10$^{-13}$ & 20.9 & 5.7$\times$10$^{-14}$ & 12.3 & 3.2$\times$10$^{-14}$ &  3$\sigma$ & 7.3$\times$10$^{-14}$ & 9.7 &  \\
DK Tau & 2.9$\times$10$^{-14}$ & 3.1 & 1.1$\times$10$^{-13}$ &  3$\sigma$ & 5.0$\times$10$^{-14}$ &  3$\sigma$ & 2.1$\times$10$^{-14}$ &  3$\sigma$ & 3.5$\times$10$^{-14}$ &  3$\sigma$ & 1.2$\times$10$^{-13}$ & 11.5 &  \\
DL Tau & 1.5$\times$10$^{-13}$ & 15.9 & 8.6$\times$10$^{-14}$ & 8.3 & 3.0$\times$10$^{-13}$ & 30.1 & 1.2$\times$10$^{-13}$ & 30.6 & 6.0$\times$10$^{-14}$ & 9.1 & 5.1$\times$10$^{-13}$ & 86.6 &  \\
DO Tau & 4.7$\times$10$^{-13}$ & 25 & 1.2$\times$10$^{-13}$ & 6.5 & 4.3$\times$10$^{-13}$ & 16.6 & 1.2$\times$10$^{-13}$ & 24.3 & 4.7$\times$10$^{-14}$ & 7.8 & 4.6$\times$10$^{-13}$ & 79 &  \\
DR Tau & 9.0$\times$10$^{-13}$ & 43 & 3.4$\times$10$^{-13}$ & 6.6 & 1.1$\times$10$^{-12}$ & 33 & 3.2$\times$10$^{-13}$ & 52.8 & 1.4$\times$10$^{-13}$ & 4 & 8.3$\times$10$^{-13}$ & 96.4 &  \\
DS Tau & 1.7$\times$10$^{-13}$ & 3.7 & 1.1$\times$10$^{-13}$ &  3$\sigma$ & 1.2$\times$10$^{-13}$ & 5.4 & 5.7$\times$10$^{-14}$ & 5.8 & 1.9$\times$10$^{-14}$ &  3$\sigma$ & 1.2$\times$10$^{-13}$ & 20 &  \\
FQ Tau & 6.9$\times$10$^{-14}$ & 20.2 & 1.3$\times$10$^{-14}$ & 3.7 & 4.6$\times$10$^{-14}$ & 10.4 & 1.8$\times$10$^{-14}$ & 11.7 & 8.0$\times$10$^{-15}$ &  3$\sigma$ & 3.0$\times$10$^{-14}$ & 9.1 &  \\
FS Tau & 2.9$\times$10$^{-13}$ & 4.7 & 1.8$\times$10$^{-13}$ &  3$\sigma$ & 2.0$\times$10$^{-13}$ & 8.5 & 5.1$\times$10$^{-14}$ & 10.6 & 1.8$\times$10$^{-14}$ &  3$\sigma$ & 9.1$\times$10$^{-14}$ & 13 &  \\
FT Tau & 2.7$\times$10$^{-14}$ &  3$\sigma$ & 2.3$\times$10$^{-14}$ &  3$\sigma$ & 1.7$\times$10$^{-14}$ &  3$\sigma$ & 1.0$\times$10$^{-14}$ & 3.5 & 1.0$\times$10$^{-14}$ &  3$\sigma$ & 1.5$\times$10$^{-13}$ & 40 &  \\
FX Tau & 2.7$\times$10$^{-13}$ & 15.9 & 7.8$\times$10$^{-14}$ &  3$\sigma$ & 1.5$\times$10$^{-13}$ & 9 & 3.8$\times$10$^{-14}$ & 3 & 3.2$\times$10$^{-14}$ &  3$\sigma$ & 3.3$\times$10$^{-14}$ & 6.7 &  \\
FZ Tau & 1.9$\times$10$^{-12}$ & 5.6 & 6.5$\times$10$^{-13}$ & 5 & 1.9$\times$10$^{-12}$ & 27.9 & 5.4$\times$10$^{-13}$ & 26.2 & 2.1$\times$10$^{-13}$ & 9.6 & 1.0$\times$10$^{-12}$ & 82.2 &  \\
GI Tau & 7.2$\times$10$^{-14}$ & 4 & 4.8$\times$10$^{-14}$ &  3$\sigma$ & 3.6$\times$10$^{-14}$ & 3.5 & 1.8$\times$10$^{-14}$ & 6.2 & 3.2$\times$10$^{-14}$ &  3$\sigma$ & 2.9$\times$10$^{-13}$ & 30.5 &  \\
GK Tau & 1.8$\times$10$^{-13}$ & 6.1 & 6.0$\times$10$^{-14}$ &  3$\sigma$ & 1.3$\times$10$^{-13}$ & 7.5 & 6.5$\times$10$^{-14}$ & 9.7 & 1.9$\times$10$^{-14}$ &  3$\sigma$ & 3.8$\times$10$^{-14}$ & 4.2 &  \\
GN Tau & 2.9$\times$10$^{-13}$ & 9.2 & 7.9$\times$10$^{-14}$ & 4.3 & 2.5$\times$10$^{-13}$ & 14.1 & 9.1$\times$10$^{-14}$ & 16.2 & 4.7$\times$10$^{-14}$ & 7.2 & 2.6$\times$10$^{-13}$ & 55.2 &  \\
Haro 6-28 & 3.9$\times$10$^{-14}$ & 6.3 & 9.9$\times$10$^{-15}$ &  3$\sigma$ & 2.4$\times$10$^{-14}$ & 7.1 & 1.1$\times$10$^{-14}$ & 9.2 & 5.7$\times$10$^{-15}$ &  3$\sigma$ & 1.6$\times$10$^{-14}$ & 16.5 &  \\
HP Tau & 6.0$\times$10$^{-13}$ & 44.4 & 1.4$\times$10$^{-13}$ &  3$\sigma$ & 8.3$\times$10$^{-14}$ &  3$\sigma$ & 7.8$\times$10$^{-14}$ & 9.1 & 4.8$\times$10$^{-14}$ & 8.3 & ... & ... & Br$\gamma$ in absorption \\
HP Tau G3 & 2.1$\times$10$^{-13}$ & 5.6 & 6.2$\times$10$^{-14}$ &  3$\sigma$ & 1.0$\times$10$^{-13}$ & 9.5 & 3.4$\times$10$^{-14}$ & 6 & 9.4$\times$10$^{-15}$ &  3$\sigma$ & ... &  ... & Br$\gamma$ in absorption \\
HQ Tau & 3.1$\times$10$^{-13}$ &  3$\sigma$ & 2.1$\times$10$^{-13}$ &  3$\sigma$ & 1.3$\times$10$^{-13}$ &  3$\sigma$ & 5.6$\times$10$^{-14}$ &  3$\sigma$ & 8.4$\times$10$^{-14}$ &  3$\sigma$ & 1.6$\times$10$^{-13}$ & 5.3 &  \\
IQ Tau & 3.3$\times$10$^{-14}$ & 3.7 & 4.5$\times$10$^{-14}$ &  3$\sigma$ & 4.6$\times$10$^{-14}$ &  3$\sigma$ & 2.0$\times$10$^{-14}$ &  3$\sigma$ & 1.5$\times$10$^{-14}$ &  3$\sigma$ & 2.7$\times$10$^{-14}$ & 4.4 &  \\
IRAS 04125+2902 & 1.1$\times$10$^{-13}$ &  3$\sigma$ & 5.9$\times$10$^{-14}$ &  3$\sigma$ & 7.7$\times$10$^{-14}$ & 7 & 3.0$\times$10$^{-14}$ & 5.9 & 1.7$\times$10$^{-14}$ &  3$\sigma$ & 2.2$\times$10$^{-14}$ & 6.6 &  \\
IRAS 04303+2240 & 9.8$\times$10$^{-13}$ &  3$\sigma$ & 4.3$\times$10$^{-13}$ &  3$\sigma$ & 1.6$\times$10$^{-13}$ &  3$\sigma$ & 6.1$\times$10$^{-14}$ & 4.4 & 4.8$\times$10$^{-14}$ &  3$\sigma$ & 2.4$\times$10$^{-13}$ & 22 &  \\
RY Tau & 5.1$\times$10$^{-13}$ &  3$\sigma$ & 2.0$\times$10$^{-13}$ &  3$\sigma$ & 2.0$\times$10$^{-13}$ &  3$\sigma$ & 9.8$\times$10$^{-14}$ &  3$\sigma$ & 1.2$\times$10$^{-13}$ &  3$\sigma$ & 1.0$\times$10$^{-12}$ & 22.6 & C$^0$ in  absorption \\
\\
\hline                                   
\end{tabular}
\\
\textbf{Notes:} Integrated fluxes for major line complexes of C$^0$ and H$^0$ Br$\gamma$ observed in this sample; in most cases these include multi-line blends. Limits of integration are as follows: 0.906-0.912 $\mu$m, 0.9395-0.942 $\mu$m, 1.0675-1.0705 $\mu$m, 1.174-1.177 $\mu$m, 1.4530-1.4560 $\mu$m, and 2.158-2.174 $\mu$m, respectively.
\end{table*}

\section{Model analysis}
\label{mod_nete}
To test whether the lines originate in the inner gaseous disk, one must determine how to produce a reservoir of neutral, atomic carbon and how to excite these particular lines. As a first order approximation to the geometrically thin midplane layer in the inner disk, I model a simple slab using a combination of a photoionization code and atomic database, as described below. 

\subsection{Carbon ionization fraction via Cloudy models}
\label{cloudy}

To test what fraction of the carbon in the inner disk should be neutral atomic carbon, C$^0$, rather than ionized carbon or CO, I used version 17.02 of the photoionization code Cloudy \citep{ferland2017} to compute the expected carbon ionization fraction for two sets of slabs: one with a vertical power law in the hydrogen density as a function of distance from the star that roughly approximated the gaussian density distribution in height expected from a disk in hydrostatic equilibrium, and one with a radially dependent hydrogen density with power -2. The midplane values in each of these slabs at 0.03 AU ranged in a grid from 10$^{14}<n_H<10^{17}$ cm$^{-3}$, the range of midplane densities within 0.1 AU in the output of the full disk \citet{dalessio+06} models used by \citet{mcclure+16}. For the radial slab, I set a stopping criterion of 0.15 AU, as an approximate location of the dust sublimation radius in typical TTS systems \citep{mcclure+13b}, and for the vertical slab I set a stopping depth that would slightly overshoot the midplane location at 0.03 AU. Since the region of interest is inside the dust sublimation radius, I did not include a dust contribution in either slab. The physics and chemistry of molecular hydrogen is important for this region, so I enabled the larger H$_2$ molecule module in Cloudy \citep{shaw2005}. 

The radiation field was a combination of ATLAS9 stellar models with solar metallicity \citep{castelli2004} at the stellar effective temperature, $T_{eff}$, and the stellar accretion shock temperature of 8000 K, taking the stellar and accretion luminosities found in Section \ref{stellparm}. I also included a contribution to the radiation field from X-rays, assuming a 10 MK bremsstrahlung emission spectrum, with a luminosity of $L_X$=10$^{30}$ erg/s between 0.3 keV and 10 keV \citep{rab2017}. For the vertical slab, I multiplied the incident flux at the top of the slab by the cosine of the incident angle to the slab normal. I included accretion heating by implementing Cloudy's $\alpha$ disk option using the stellar mass found in Section \ref{stellparm}. 

\begin{figure*}
   \label{cloudies}
   \centering
   \includegraphics[width=\hsize]{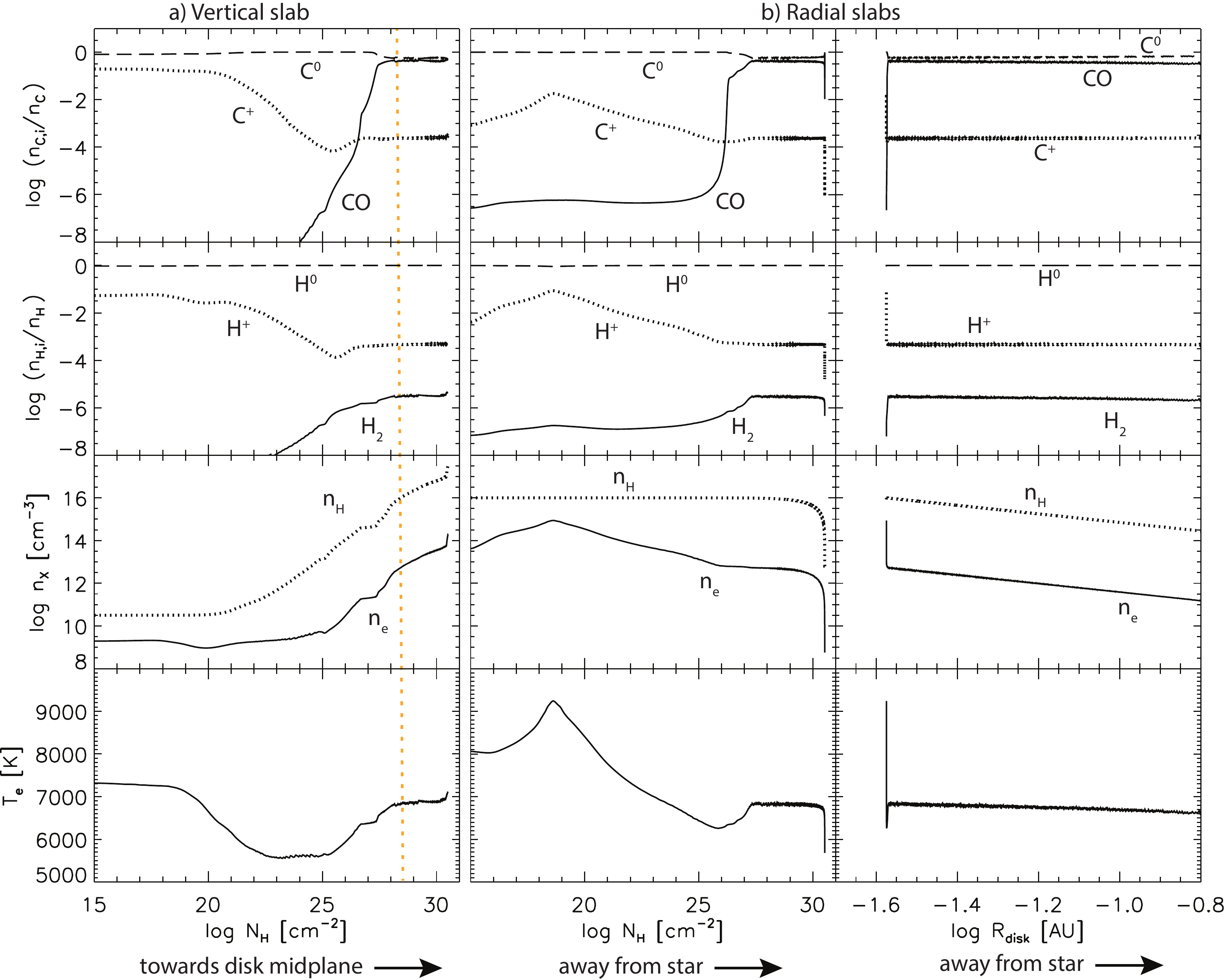}
      \caption{Two 1D Cloudy slab models of the inner disk around $\sim$0.03 AU. Input stellar parameters are for DR Tau. {\it Left:} Physical properties are plotted as a function of hydrogen column density for a vertical slab located at 0.03 AU with a power law density structure extending towards the disk midplane. These include: fractional abundance of dominant ionization stages for carbon (top panel) and hydrogen (second panel), hydrogen and electron densities (third panel), and electron temperature (bottom panel). The dashed vertical orange line indicates the location where the slab reaches n$_H$=10$^{16}$ cm$^{-3}$. {\it Middle:} An analogous plot, but for a radial slab of initial n$_H$=10$^{16}$ cm$^{-3}$ at the co-rotational radius and a power law dependence on disk radius extending away from the star. {\it Right:} The same as the middle plot, but plotted against the logarithm of the disk radius instead of N$_H$. Most of the very high density and temperature structure seen in the N$_H$ plot occurs at the corotational radius due to the high optical depth.}
   \end{figure*}

\begin{figure}
   \label{nenh}
   \centering
   \includegraphics[width=\hsize]{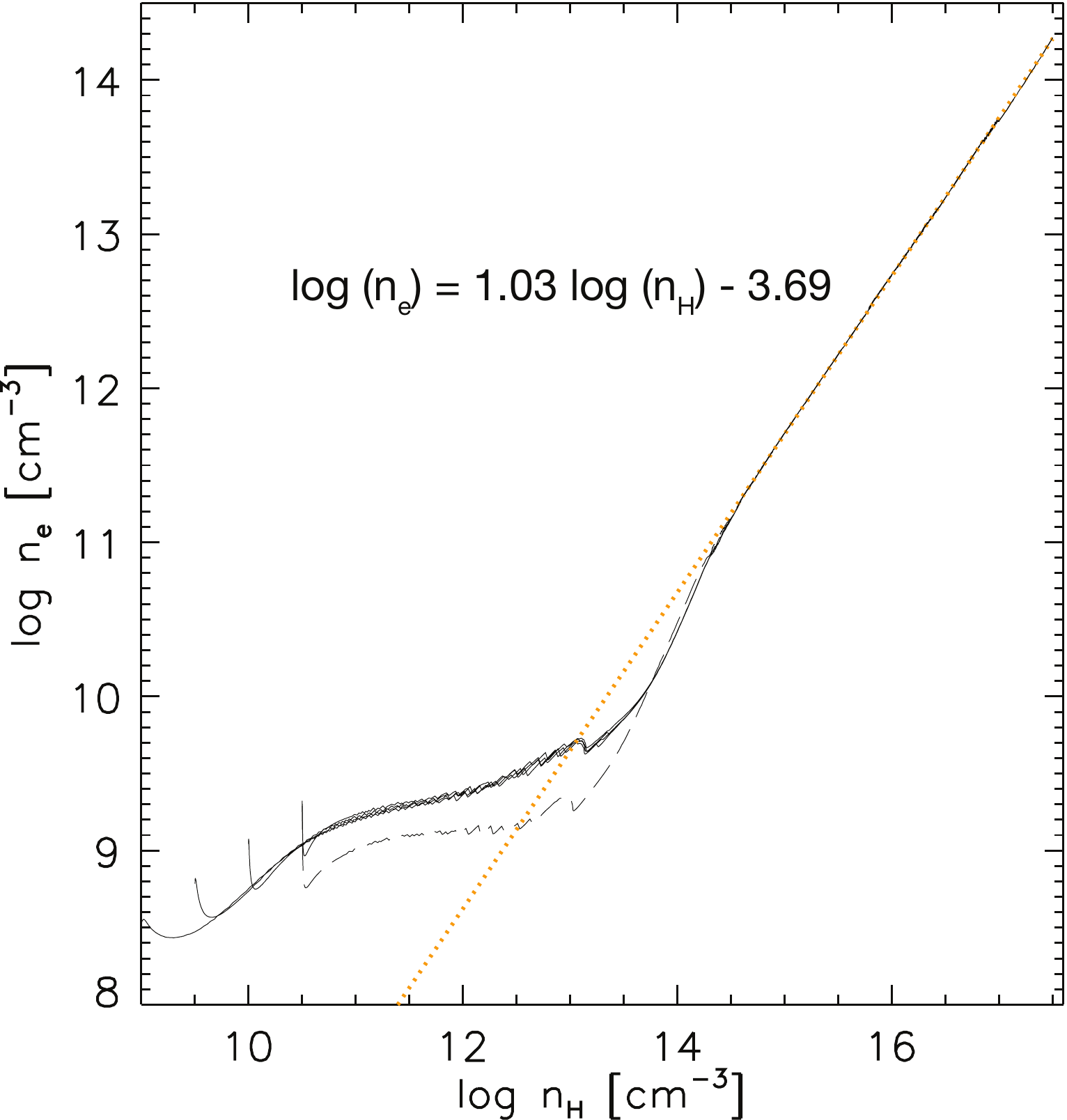}
   \caption{Electron density versus hydrogen density for vertical slab models of stars in my sample. The orange dotted line represents the power law fit to the region with log $n_H$ > 14.5, and labelled equations gives the functional form of this fit.}
   \end{figure}

The resulting physical conditions are shown in Fig. \ref{cloudies} for the stellar parameters of DR Tau. In the vertical slab (left panel), with a starting density of $n_H=10^{10.5}$, the carbon content of the slab is predominantly neutral at densities less than $n_H\sim10^{15}$ cm$^{-3}$.   At the midplane, with hydrogen densities $n_H\sim10^{16}$ cm$^{-3}$, the fractional abundance of carbon in CO increases to a limiting value of $\sim$0.5; this limit results from collisional dissociation of CO at such high densities. {\it So conservatively, C$^0$ carries at least 50\% of the bulk carbon at the midplane.} The majority of hydrogen is also in neutral atomic form, with H$^+$ making up the next largest contribution. Interestingly, at the midplane density, $n_H\sim10^{16}$ cm$^{-3}$, the fractional molecular hydrogen plateaus around 10$^{-6}$, while the ionized hydrogen fraction remains at 10$^{-3.5}$, resulting in an electron density larger than $n_e\sim10^{13}$. For all of the stars in my sample, the relationship between $n_e$ and $n_H$ was a power law form for log($n_H$)$\ge$14.5, as seen in Figure \ref{nenh}: $log(n_e)=1.03 log(n_H) - 3.69$.

In the radial slab model (Fig. \ref{cloudies}, right panel), the temperature and electron density peak at a column density of $N_H=10^{19}$ cm$^{-2}$, where most of the stellar radiation is deposited. At midplane densities greater than  $n_H\sim10^{15}$ cm$^{-3}$ the slab quickly becomes optically thick to the stellar radiation in the radial direction, reaching an $N_H\sim$10$^{25}$ at R=3.008 $R_{*}$. At this column density, even the x-rays in the model are attenuated. Consequently the slab has a short, illuminated `face' that is both radiatively and collisionally excited as well as a larger segment for which the ionization of H$^0$ is dominated by charge exchange collisions. The electron temperature exhibits an inversion, with a minimum near $N_H\sim$10$^{25}$ cm$^{-2}$ and a plateau of $\sim$7000 K at $N_H\sim$10$^{28}$ cm$^{-2}$. As with the vertical slab, the C$^0$ fraction is always at least 50\% of the total carbon population.

The causes and implications of the high $T_e$ and $n_e$ are discussed in Section \ref{chem}. I carry forward the minimum {\it C$^0$ fraction (>50\%), range of atmosphere and midplane $T_e$ and $n_e$, and the power law dependence of $n_e$ on $n_H$} for the midplane in order to model the C$^0$ line emission in the following section.

\subsection{Electron densities and temperatures}
Under the assumption that the emission arises in an optically thin region, which I verify a posteriori in Appendix \ref{optdepthapp}, the line emission depends on the electron density, $n_e$, and temperature, $T_e$, as well as the amount of emitting material, $n_{C^0}$. The latter depends on the abundance of carbon with respect to hydrogen, $X_C$, the fraction of carbon in C$^0$, $n_{C^0}$/$n_C$, and the size of the emitting region. By taking flux ratios between different C$^0$ emission lines, one can constrain $n_e$ and $T_e$ independent of $X_C$ or $n_{C^0}$/$n_C$, assuming that both lines originate in the same region. Using the atomic data and routines provided by the version 8.0.2 of the CHIANTI atomic database \citep[][and references therein]{dere+97,landi+13} to calculate the level populations, I determined the line intensity produced by a slab extending from the approximate co-rotational radius at 3 stellar radii, R$_{*}$, to an outer radius, R$_{out}$, defined below: 

\begin{equation}
I_{\lambda_{ij}}=\frac{h\nu_{ij}}{4\pi}\int n_jA_{ji}dr = \frac{h\nu_{ij}}{4\pi} \frac{A_{ji}}{n_e}X_C\frac{n_{C^{0}}}{n_C}\frac{n_{j,C^{0}}}{n_{C^{0}}}\int_{3R_*}^{R_{out}} n_Hn_edr \\
\end{equation}

The outer radius of the composite slab is taken from literature estimates based on kinematics of the CO fundamental or H$^0$ Br$\gamma$ interferometry, as described in Appendix \ref{outradapp}. The effect of optical depth to the stellar radiation is approximated by taking a two-component slab with the same electron density and temperature in each component, but with photoexcitation allowed only interior to 3.008R$_*$. To find the best-fitting $n_e$ and $T_e$ independent of emitting area or hydrogen density, there are line pairs that are excellent measures of the density (1.1757/1.0694 $\mu$m and 1.0686/1.0694 $\mu$m) and temperature (0.9408/1.0694 $\mu$m). However, the strongest lines in my sample at $\sim$1.07 and $\sim$1.175 $\mu$m are three-line blends that are not spectrally resolved here. Therefore, rather than constraining $n_e$ and $T_e$ from integrated line fluxes, I normalized the continuum subtracted excess spectra to the flux at 1.0694 $\mu$m, and performed a least-squares fit of the data to a similarly normalized grid of models.  

\begin{figure*}
   \centering
   \includegraphics[width=21cm]{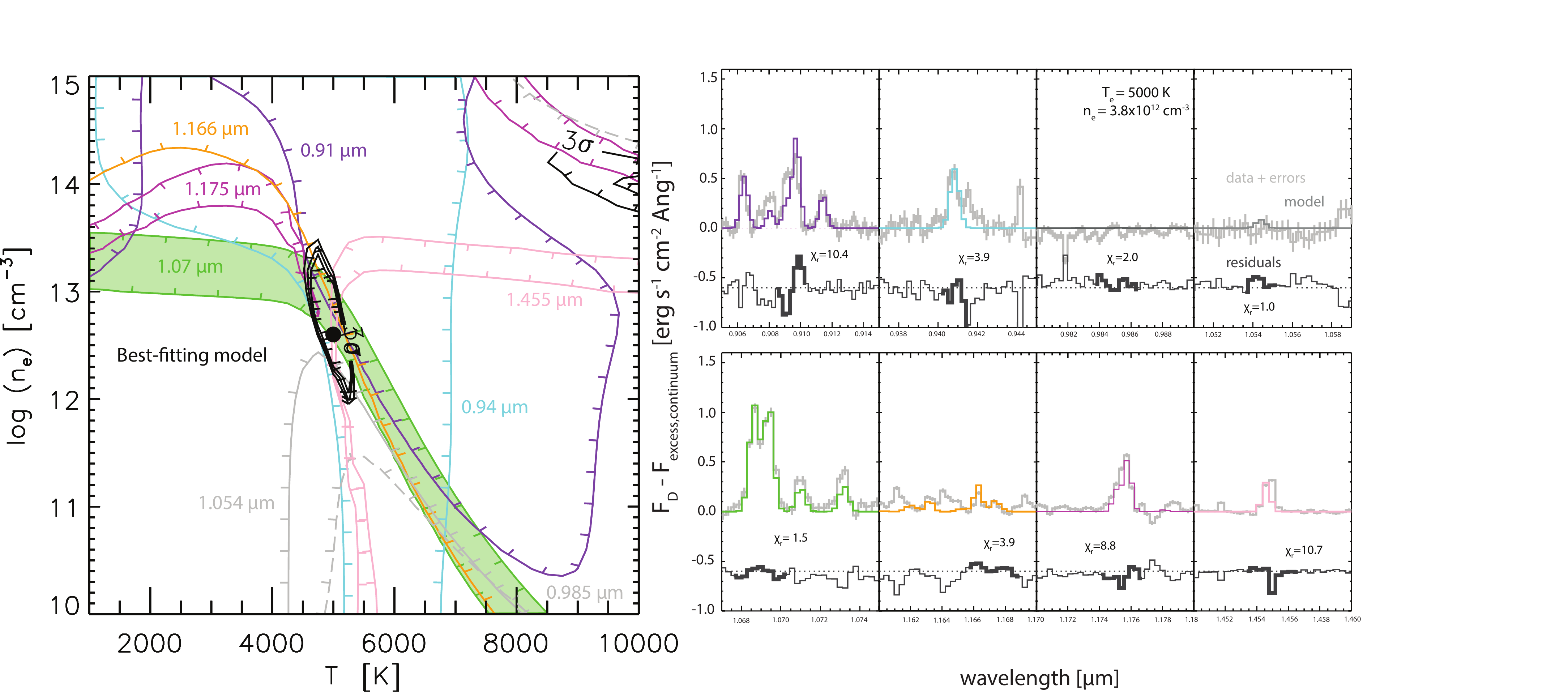}
   \caption{Left: The electron temperatures and densities that fit the ratios between the 1.069 $\mu$m line and all other lines at a 1$\sigma$ level (colored contours, wavelengths labeled). The lines are color coded to match the spectra in the right panel. The green fill in between the green contours is an example of how to read the contour plot; the filled region covers the best-fitting set of $n_e$ and $T_e$ for the 1.07 $\mu$m line ratio. Other contours are not filled, but the tickmarks point downhill. The gray contour indicates where the non-detection at 1.0544 $\mu$m is matched. The black contours are the 1, 2, and 3$\sigma$ confidence intervals on the best-fit to the combined lines. Right: The continuum subtracted spectra of the six main C$^0$ transition clusters and the non-detections at 0.985 and 1.0544 $\mu$m, normalized to the flux at 1.0694 $\mu$m, in DO Tau. Model fits for T$_e$=5750 K and n$_e$=6.3$\times$10$^{12}$ cm$^{-3}$ are overplotted. The spectra are color-coded to match the 1$\sigma$ line ratios regions in the left panel. }
              \label{tenes}%
    \end{figure*}

\begin{figure*}
   \centering
   \includegraphics[width=17cm]{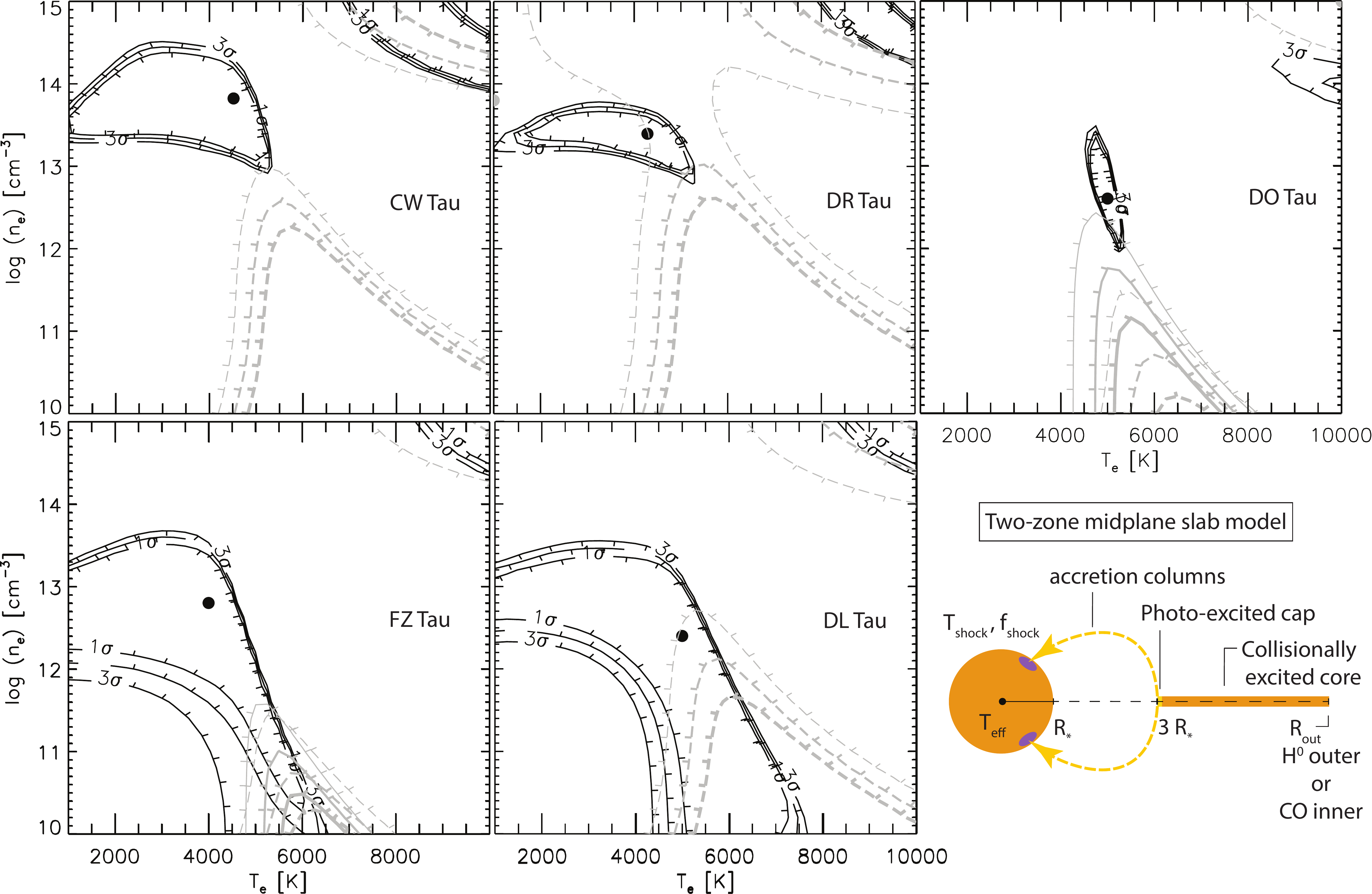}
   \caption{Grids of $n_e$ and $T_e$ for the C$^0$ line ratios of each of the five stars. Black contours are the 1, 2, and 3$\sigma$ confidence intervals on the best-fitting combination of $n_e$ and $T_e$ for the detected emission lines, given by the black dot. Gray contours are the 1, 2, and 3$\sigma$ confidence intervals on the non-detected lines at 0.985 and 1.055 $\mu$m. The non-detections preclude the high density, high temperature solutions to the detected line fits seen in, e.g., CW Tau. The bottom right panel is a representative cartoon of the slab model, with the photoexcited cap and optically thick, collisionally excited core.}
   \label{alldensity}%
              
    \end{figure*}

Figure \ref{tenes} shows the 1$\sigma$ contours for the best-fitting range of $T_e$ and $n_e$ for all lines in one disk, DO Tau. The main lines at 1.07$\mu$m are well-fit by a wide range of parameter space, spanning high densities and low temperatures ($\sim10^{13}$ cm$^{-3}$ and 3000 K) to lower densities and higher temperatures ($\sim10^{11}$ cm$^{-3}$ and 7000 K). However, the next strongest lines at 1.175 $\mu$m are only fit by high densities and lower temperatures, while the remaining lines are better fit by lower densities and higher temperatures. In all cases, it is possible to produce a `best fit' that does a reasonable job of fitting most of the lines at an intermediate temperature and density. Figure \ref{alldensity} shows the confidence intervals for the best-fitting slab. Generally, the electron densities are all high, indicating that the lines must arise in the disk midplane layer. 

\subsection{Carbon abundances and uncertainties}
\label{coh}
Using the best-fit $T_e$ and $n_e$, I fit the absolute flux of each line cluster individually to determine a carbon density.  The mean and uncertainty of these carbon densities for the combined line set, $\overline{n_C}$, are listed in Table \ref{carbdep}. It is then possible to use the relationship between the midplane $n_H$ and $n_e$ found in the Cloudy models of Section \ref{cloudy} to compute the $n_H$ implied by these $n_e$ values. I combine this with the observed $n_C$ to obtain $X_C$, also listed in Table \ref{carbdep}. The gas phase carbon abundances are all less than solar, by up to a depletion factor of $\sim$42. I discuss the implications of this result in Section \ref{cdep}. 

The abundances and depletion factors in Table \ref{carbdep} are reported with error bars propagated from the 1$\sigma$ uncertainties on $\overline{n_C}$. Within the uncertainties two disks, DL Tau and FZ Tau, have carbon abundances that are consistent with solar values, while three others, DR Tau, DO Tau, and CW Tau, have robust carbon depletion.

In addition to the scatter in measurements of $n_C$ for each disk, there are three sources of systematic uncertainty in the C/H values found with my method. The first is in the derivation of $n_H$ from the relationship with $n_e$ in the Cloudy slab models.  At these high densities, the free electrons are generated mainly through charge exchange reactions, particularly with sulphur-bearing ions \citep{kingdon1999}. The reaction coefficients are accurate to within a factor of 2 to 4 \citep{kingdon1996}, an uncertainty which carries over into the C/H measurements. A second source of uncertainty is the C/O ratio of the gas. Solar C/O ratios produce CO/C$_{total}$ ratios of at most 0.5, which is what I have taken into account in the final $n_C/n_H$ calculation. This upper limit to CO/C holds down to at least C/O$\sim$0.1; very low C/O values are unlikely, since the various mechanisms to deplete carbon would also deplete oxygen. One way to resolve this issue is to include fits to the flux of the CO 4.7 $\mu$m fundamental line in future numerical simulation efforts; this would directly measure CO/C$^0$, particularly if combined with kinematic estimates of the emitting region size (see below).

A third source of uncertainty is the emitting volume, which is used to fit $\overline{n_C}$; the majority of the uncertainty is in the radial extension of the region. Although the vertical structure of the inner disk is important to understand the propagation of stellar radiation, for this sample of disks the C$^0$ is well-constrained to originate in a narrow vertical region around the disk midplane, which is much smaller than the height of the dust sublimation rim \citep{muzerolle04}. This is because the $n_e$ found from the C$^0$ line ratios can only be explained by $n_H$ values so large that the gas has become optically thick to the stellar radiation, which only happens close to the midplane (assuming that the disk is vertically in hydrostatic equilibrium). Therefore the uncertainty in emitting volume can be reduced to the radial extent of the slab. In this case, a smaller emitting region would drive $\overline{n_C}$ to higher values and therefore to a lower degree of carbon depletion. The radial location of the dust sublimation rim in the disk midplane is a hard upper limit for the outer edge of the emitting region, as it is only possible to prevent the reformation of H$_2$ from H$^0$ in the absence of dust grains. To better constrain the outer radius of the emitting region, R$_{out}$, future studies should combine kinematically determined {\it outer} radii of the C$^0$ emission itself with interferometrically determined {\it outer} radii of Br $\gamma$ emission and kinematically determined {\it inner} radii of CO from the 4.7$\mu$m fundamental line. According to the Cloudy slab models, these measurements should identify where the disk midplane transitions from mostly molecular to mostly atomic, and therefore where the C$^0$ emission can originate.

The latter two uncertainties will be easily resolved with future observations using recently commissioned or upcoming instruments, such as the near- and mid-infrared interferometers VLT-PIONIER and -MATISSE, and high-resolution infrared spectrometers IRTF iSHELL and VLT CRIRES+ to kinematically constrain the emitting region. JWST NIRSPEC will be able to detect these lines at medium resolution even in samples of very faint disks. Additionally, both JWST NIRSPEC and MIRI observations of the molecular gas outside of the atomic zone will provide robust column densities of all isotopologues of the main carbon-bearing molecular species, which will provide a lower-limit check on the total carbon column densities found in the atomic line region. Improvements to the current Cloudy simulations to make them 1+1D would demonstrate better the impact of the vertical disk structure and stellar radiation field, particularly for disks with lower values of $n_H$, in which photoprocesses play a larger role in the line emission. Systematic uncertainties on the $n_H$ measurements would be improved by calculating or measuring more accurate charge exchange rates, and by separating out the H$^0$ contribution from the accretion shocks from that of the disk. Finally, benchmarking my simulations with other disk chemical models (e.g. DIANA or DALI) using different reaction networks would also help to understand the impact of these systematic uncertainties on the accuracy of the C/H values calculated here.

\begin{table*}
\caption{Densities and depletion factors}             
\label{table:3}      
\centering                          
\begin{tabular}{c c c c c c c c c c}        
\hline\hline                 
\\
Star  & R$_{out}$ & log($n_e$) & $T_e$ & log($\overline{n_C}$) & log($n_H$) & C/H  & Depletion factor & $\Sigma_{H, 0.03 AU}$ &Missing C\\    
    &     [AU] & [cm$^{-3}$] & [K]  & [cm$^{-3}$] & [cm$^{-3}$] & [solar] & -  & [g cm$^{-2}$] & [$M_{\oplus}$] \\
\hline                        
CW Tau & 0.04 & 13.8 &  4500 &  11.5 (0.1) &    17.0 &    2.4$^{+0.6}_{-0.5}\times$10$^{-2}$ & 42$^{+10}_{-9}$ & 4142 & 2.9$\times10^{-3}$\\
DR Tau & 0.05 & 13.4 &  4250 &  11.7 (0.2) &    16.7 &    7.4$^{+4.4}_{-2.7}\times$10$^{-2}$ & 13$^{+8}_{-4}$ & 2792 & 2.5$\times10^{-3}$  \\
FZ Tau & 0.1 & 12.8 &  4000 &  11.3 (0.3) &    16.1 &   1.2$^{+1.2}_{-0.6}\times$10$^{-1}$ & 8$^{+9}_{-4}$ & 678 & 3.6$\times10^{-3}$ \\
DO Tau & 0.1 & 12.6 &  5000 &  10.7 (0.3) &    15.9 &   4.6$^{+4.8}_{-2.2}\times$10$^{-2}$ & 22$^{+20}_{-11}$ & 476 & 2.1$\times10^{-3}$ \\ 
DL Tau & 0.05 & 12.4 &  5000 &  11.1 (0.2) &    15.7 &    1.9$^{+1.0}_{-0.7}\times$10$^{-1}$  & 5$^{+3}_{-2}$ & 252 & 1.9$\times10^{-4}$ \\
\\
\label{carbdep}
\end{tabular}
\\
\textbf{Notes:} Results of the CHIANTI slab model fits to the 1) C$^0$ line ratios for $n_e$ and $T_e$ and 2) absolute line fluxes for $n_C$. The values of $n_H$ are calculated from the $n_e(n_H)$ dependence from the Cloudy slab models. All disks show a depletion in the C/H abundance relative to the solar value, which is taken to be 2.69$\times$10$^{-4}$. The surface density of hydrogen gas was calculated assuming a density profile in hydrostatic equilibrium with a midplane temperature equal to $T_e$. The mass of `missing' carbon inferred for a static disk from the depletion factor is given in the last column. Notes: References for outer radii are given in Appendix B.
\end{table*}

\section{Discussion}
\label{discussion}
The detection of hot, dense gas inside the dust sublimation rim has important implications for several open topics in the disks and planet formation community, namely 1) the results of grain surface chemistry in the outer disk, 2) observational constraints on the efficiency and composition of planetesimals formed by $\sim$2 Myr, and 3) the structure and physical properties of the little-explored inner gaseous disk.

\subsection{Implications of measured carbon abundance}
\subsubsection{Confirmation of carbon depletion, rather than lower mass}
\label{cdep}
This work is the first evidence for a gas phase carbon depletion, relative to the solar value of C/H=2.69$\times$10$^{-4}$, inside the dust sublimation radius of protoplanetary disks. It is also noteworthy that this carbon abundance is inferred relative to hydrogen, via the electron densities, so there is no question of a major degeneracy between carbon abundance and hydrogen mass. Previous work using chemical models to analyze far-infrared and submillimeter HD, CO, and C$^0$ line emission also found that carbon is depleted from the gas phase in the outer disk of TW Hya \citep{bergin+13, schwarz+16a, kama+16b}, and potentially two younger disks, DM Tau and GM Aur \citep{mcclure+16}. There has been discussion of whether these results are more widely applicable to other systems, given the degeneracy between C/H and the dust/gas ratio in interpreting ALMA observations using chemical models \citep{miotello2017}. The detection of carbon depletion here in a different sample using independent wavelengths, techniques, and disk radii gives greater credence to the previous results and strengthens the case that, in some systems, a lower gas mass as derived from CO represents a change in the gas phase carbon abundance rather than a lower H$_2$ mass per se. 

\subsubsection{Causal mechanisms for C depletion}
\label{mech}
\begin{figure*}
   \centering
   \includegraphics[width=15cm]{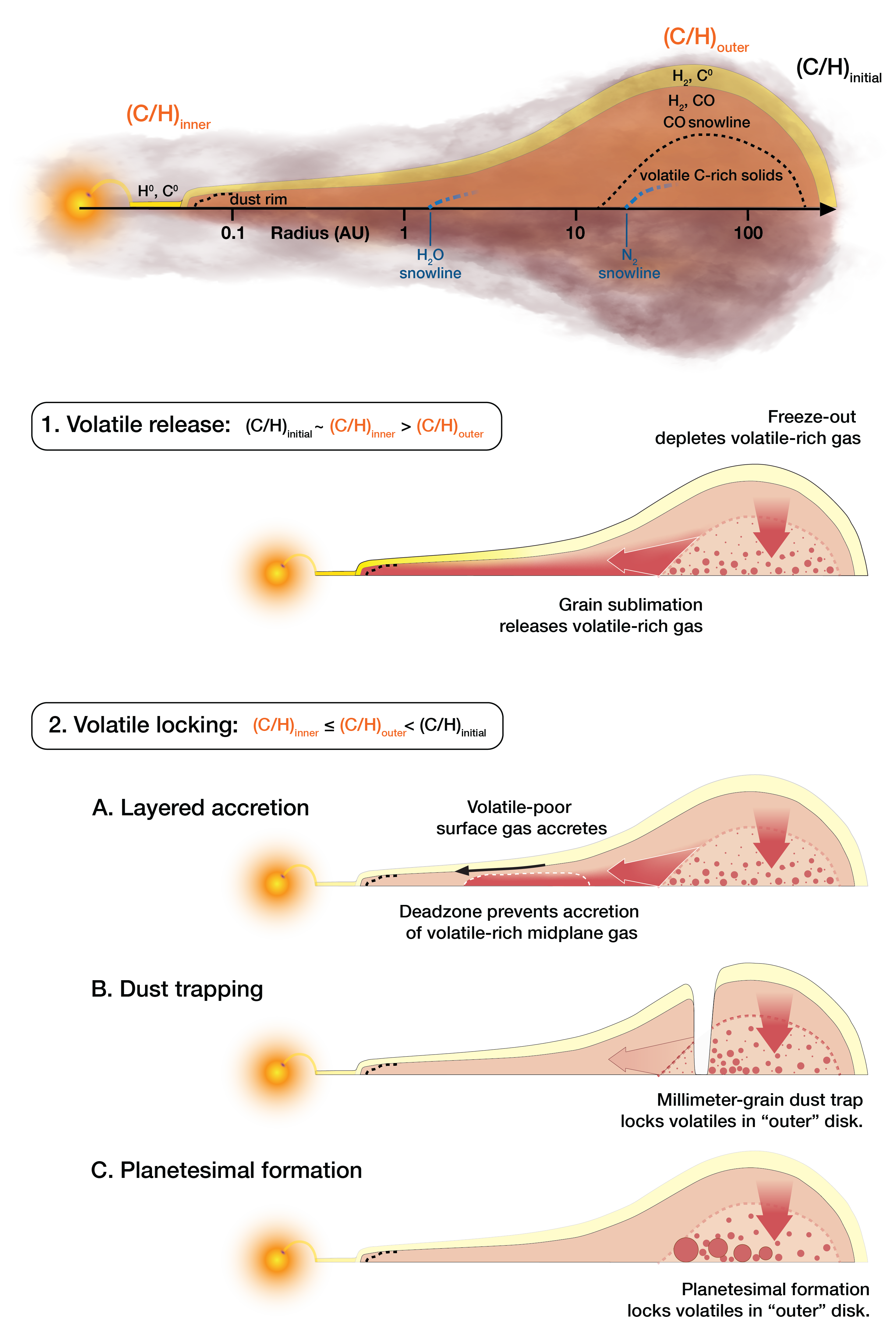}
   \caption{Top: schematic of the physical locations in the disk corresponding to different carbon carriers (dust, molecular gas, and atomic gas), with labels for terms used in sketches below. Notional locations of midplane water ice snowline and N$_2$ ice snowline are indicated by dashed blue lines that, for clarity, fade out with radius. Middle: illustration of the carbon release scenario, which returns the inner disk gas carbon abundance to the original value. Bottom: three options for the carbon locking scenario: a) a deadzone blocking accretion through the midplane, and b) dust traps and c) planetesimal formation preventing the radial drift of volatile-rich dust grains. See text in Section \ref{mech} for more information.}
   \label{mechoptions}%
    \end{figure*}

The outer disk carbon depletion found with ALMA could be accomplished by freeze-out of CO and hydrocarbons onto large dust grains \citep{yu2017,favre+13}. This `missing' carbon should be `released' to the gas phase when the dust grains interior to the snowline corresponding to the sublimation temperature for the solid carbon carrier. Depending on the size of the grains, they may either move approximately with the gas at the disk accretion rate or by radially drifting in faster than the gas accretes. In both cases, the gas measured in the inner disk would have returned to at least the initial C/H abundance (see Fig. \ref{mechoptions}, middle panel). For the gas in the inner disk to be depleted of carbon, one must avoid accreting the volatile-rich midplane gas resulting from the sublimation of the icy grains. This can be accomplished either by preventing the gas accretion or by preventing the grains from sublimating, e.g. stopping their radial drift (Fig. \ref{mechoptions}, bottom panel, options B and C).

Midplane volatile C gas accretion can be circumvented by the presence of a deadzone in the midplane, at radii and depths where the disk is not longer thermally or X-ray ionized \citep{gammie1996}. Volatile-poor gas from the outer disk could then be accreted through the ionized surface layer, while the volatile-rich gas is sequestered in the deadzone. However, more than 50\% of the total ISM carbon budget is thought to be contained in refractory-rich volatile grains, which sublimate at temperatures where the MRI should already be active at the disk midplane. Therefore, unless radial drift is inefficient, refractory carbon-rich dust grains would drift through the deadzone and return the inner disk gas phase content to within a factor of 2 of the initial value. In the sample of five modeled disks, four have measured outer dust radii that are substantially less than their outer gas radii \citep[][and van Dishoeck, private communication, for DR Tau]{najita_bergin2018}, with the final disk, FZ Tau, lacking the gas disk outer radius measurement. The discrepancy between the dust and gas radii suggests that these disks have experienced efficient radial drift. To deplete their inner disk gas substantially of carbon, some process must be preventing the radial drift of carbon-rich dust grains. 

The formation of pressure bumps, which trap dust, or sufficiently large and fast growth can both stop grain from radially drifting. Dust traps typically produce a visible discontinuity in the millimeter grain distribution, e.g. rings or asymmetries. Of these disks, DL Tau shows clear gaps and rings in recent millimeter continuum observations \citep{long2018b}; however, it is also the disk with the least carbon depletion. In contrast, DR Tau has a smooth distribution, but an order of magnitude carbon depletion. In this disk, dust traps cannot be responsible for the observed carbon depletion, leaving grain growth to large bodies as the best explanation for carbon locking in at least a subset of disks.
%

\subsubsection{Properties of kilometer size, carbon-rich planetesimals}
\label{lock}

I estimate the size of the formed bodies by equating the drift timescale \citep{birnstiel+12}, 

\begin{equation}
\tau_D =  \frac{R_D}{u_D} = \frac{R_D V_K}{c_s^2\gamma} \Bigg(\frac{1}{St}+\frac{1}{St^{-1}} \Bigg) 
\end{equation}

\noindent to the total age of the system found in Section \ref{stellparm}. The quantities in the expression for the drift timescale are taken from the model of DO Tau published in \citet{mcclure+15} and are defined as: $R_D$, the distance between the formation region to the CO or N$_2$ snowlines at $\sim$ 30 AU, $u_D$, the drift velocity, $V_K$, the Keplerian velocity, $c_s$, the sound speed, $\gamma$, the absolute value of the power-law index of the gas pressure profile, and $St$, the Stokes number, which is dependent on the solid body size. If the large body forms at $\sim$31 AU, then it must quickly grow to 3.2 km in order to avoid drifting past the CO snowline within 2 Myr. In contrast, if it forms near 100 AU, it must only be 12.6 m in radius to remain beyond the CO snowline in that time. 

Direct comparison of the inner and outer radii carbon abundances in the same protoplantery disk would test the efficiency of the large body formation process. Unfortunately, given the high electron densities required to detect these C$^0$ lines, they would not be expected to appear in transitional or pre-transitional systems like TW Hya, DM Tau, or GM Aur, for which the outer carbon depletion factor has been estimated, so it is not yet possible to directly compare the inner and outer disk carbon depletions in the same protoplanetary disk. Future observations of the [C$^0$] line at 610 $\mu$m and CO ladder in the current sample of sources would make this comparison possible. However, with an assumption about the initial bulk composition of the gas, I can calculate a lower limit to the mass of `missing' carbon that has been locked into kilometer size bodies. Since the mass of hydrogen in the slab models used to fit the line fluxes, $M_H$, is simply $V_{slab}n_Hm_p$, then the mass of gaseous carbon is $M_{C, gas}=X_{C, dep}M_H$. If the combined gas and ice abundances were solar in the outer disks of the stars in my sample, then the minimum mass of carbon locked in planetesimals is $M_{C, solids}=12 M_H(X_{C,solar}-X_{C, dep})$. These values are listed in Table \ref{carbdep} and are typically around 3$\times$10$^{-3}$ M$_{\oplus}$. Since the disk is not static, this is obviously a gross lower limit.

\begin{table}
\caption{DR Tau inner disk gas versus planetesimal composition}             
\label{table:4}      
\centering                          
\begin{tabular}{c c c }        
\hline\hline                 
Parameters & Value & Value (solar units) \\
\hline                      
Inner gas \\
\hline                      
C/Si$^a$ & 2.72 & 3.3$\times$10$^{-1}$ \\
N/Si$^a$ & 0.20 & 9.6$\times$10$^{-2}$ \\ 
C/H &1.99$\times$10$^{-5}$ & 7.4$\times$10$^{-2}$ \\
Si/H & 7.37$\times$10$^{-6}$ & 2.3$\times$10$^{-1}$  \\
N/H & 1.49$\times$10$^{-6}$ & 2.2$\times$10$^{-2}$ \\
\hline
Planetesimals \\
\hline
C/Si (atomic) & 9.95 & 1.20 \\
N/Si (atomic) & 2.64 & 1.26 \\
N/C (atomic) & 0.27 & 1.06 \\
\hline
M$_{Si}$ (M$_{\oplus}$) & 5.8$\times$10$^{-4}$  & - \\
M$_{N}$ (M$_{\oplus}$) & 7.6$\times$10$^{-4}$ & - \\
\hline
\label{ardila}
\end{tabular}
\\
\textbf{Notes:} From \citet{asplund2009}, I take the following solar values: $(C/Si)_{\sun}$ = 8.30, $(N/Si)_{\sun}$ = 2.09, $(N/C)_{\sun}$ = 0.25, $(C/H)_{\sun}$ = 2.69$\times$10$^{-4}$, 
$(Si/H)_{\sun}$ = 3.24$\times$10$^{-5}$, $(N/H)_{\sun}$ = 6.76$\times$10$^{-5}$. References: $(a)$ \citet{ardila2013}.

\end{table}

\subsubsection{Inferred origin beyond the N$_2$ snowline}
\label{icycomp}
The carbon masses found above would imply drastically different planetesimal masses, depending on the type of planetesimal into which the carbon was locked. For example, the bulk carbon abundance in planetesimals ranges from C=31$\pm$7 wt.\% \citep[Comet 67P, ][]{fray2017} and C=3.65 wt.\% \citep[CI chondrites Orgueil/Ivuna,][]{alexander2012b} to C=0.0044 wt.\% \citep[Earth, ][]{kargel1993}. These bodies originate beyond the water ice snowline, between the snowline and the organic sootline, and between the sootline and the inner dust rim where even graphite dust will sublimate, respectively. 

The question of which solid-state species is the dominant carrier of this carbon is of interest to discussions of optimizing `stickiness' during collisional grain growth \citep{musiolik2016} and exoplanet composition \citep{madhusudhan12b}. In order to probe the composition, ideally one would measure absolute abundances of silicon as well in the gas phase.  \citet{ardila2013} derive relative luminosities of silicon, nitrogen, and carbon in the pre- and post-shock regions of the accretion streams of T Tauri stars from UV emission from high ionization state lines. Although their technique is different from mine, with a different range of temperatures, ionizations, and physical structures, it is interesting to compare the relative abundances measured for the one star in common between our samples, DR Tau. If one assumes that the variations seen in their values of $L_{Si IV}/L_{C IV}$ between objects are due to differences in the abundance of Si/C, rather than variations in the local conditions, then one can calculate the gas phase C/Si and N/Si in the accretion stream. These values are a factor of 3 and 10.5 times less than the solar C/Si and N/Si ratios, respectively (see Table \ref{table:4}). 

To determine whether this depletion arises predominately from volatile or rocky elements, I use my C/H value to calculate N/H and Si/H. Indeed, silicon is depleted from the accreting gas by a factor of 4.4, and nitrogen by a factor of 45.5, relative to the solar values. 

Combining the absolute abundance of carbon, C/H, found here with Si/C and N/C in their paper yields estimates of Si/H$_{gas}$ and N/H$_{gas}$. The amount of `missing' Si and N can be computed with the same method used above for C, resulting in masses on the order of 10$^{-4}$ M$_{\oplus}$. The number density ratios of C/Si and N/Si for the kilometer size planetesimals are then enhanced over solar values by 20\% and 26\%, respectively (Table \ref{table:4}). This result can directly tie the missing carbon and nitrogen to a particular formation region, i.e. the proto-Kuiper belt. In \citet{pontoppidan+14}, C/Si ratios for meteorites and Earth are on the order of 1 to 10$^{-3}$, respectively, both of which represent a significant solid state {\it depletion} of carbon rather than an enhancement. In contrast, comets show C/Si$\sim$10, an enhancement of carbon similar to what I find here. 

The contrast is more striking for nitrogen, which is depleted in all bodies, including the two comets in \citet{pontoppidan+14}. However, it is worth noting that the nitrogen content of planetesimals would depend strongly on whether they formed outside of the N$_2$ snowline. Molecular nitrogen is expected to be the bulk carrier of nitrogen in molecular clouds and the outer regions of disks. If embedded in a water ice matrix, the sublimation temperature of N$_2$ shifts from 18 K to 31 K, placing it at nearly the same location as the CO snowline (T$_{sub}$=34 K in water ice, vs 21 K otherwise) and the CH$_4$ snowline at 30 K, or around 30 AU, in a `typical' T Tauri disk \citep{piso2016}. Icy planetesimals formed in this region are thought to be the parent bodies of Ultracarbonaceous Antarctic Micrometeorites, which contain large fractions of irradiated N- and C-rich polyaromatic organic compounds with N/C values ranging up to solar \citep{dartois2017, engrand2018a}. The `locked' solid state values found for carbon and nitrogen in DR Tau are broadly consistent with the formation of kilometer size planetesimals in a proto-Kuiper belt region by $\sim$1 Myr, the age found for the stellar parameters derived here.

It is worth noting that for another disk in which I measure a smaller carbon depletion, DO Tau, the presence of icy planetesimals in the midplane beyond 30 AU is inferred from a tentative detection of water ice at 63 $\mu$m with the {\it Herschel} Space Observatory \citep{mcclure+15}. The crystalline nature of the ice, given the timescale for ice amorphitization at these radii, may arise through collisions between large icy bodies. Unfortunately PACS spectra for the remaining sources in my current sample are either non-existent or too noisy to be conclusive regarding the presence of ice. Further observations with SOFIA HIRMES or a future far-IR space telescope, e.g. OST or SPICA, would be illuminating. High sensitivity observations of ices in edge-on disks, through the upcoming JWST programs IceAge ERS (P.I. McClure) and MIRI GTO (P.I. van Dishoeck), will probe the nitrogen to water ratio of the ice in the upper layers of the outer disk, which will help to determine the abundance and purity of N$_2$ and hence the probable location of such kilometer size bodies.

\subsubsection{Comparison with higher mass stars}
An analogous analysis has been made for the composition of accreting gas in Herbig AeBe star systems. Due to their exterior radiative layer, which does not mix into their convective core, the photospheric abundances of these stars should reflect the composition of recently accreted gas.  \citet{kama+16b} demonstrated that Herbig stars do not show any depletion in carbon or oxygen; instead they have on average solar abundances. In contrast, many of the Herbigs with transition disks show refractory depletion in their stellar photosphere, indicating that non-icy kilometer size bodies have formed. These patterns are also broadly consistent with a scenario in which ices are the prime candidate for the carbon carrier. Since Herbig disks are warmer than T Tauri stars at large radii, due to the higher stellar radiation field, they will have a smaller region in which ices can form after infall from the molecular cloud. Additionally, their larger UV fluxes can photodesorb the ice mantles from their grains over a large radial area, leading to a lower total mass in ices. Since it is less efficient to form planetesimals via core accretion at large radii, and there may be less ice available in Herbig disks, it may be more difficult to form kilometer size planetesimals beyond the CO snowline in these disks than in T Tauri disks. Therefore, more of the carbon locked into the ices would be likely to return to the gas phase inside of the ice snowlines, and the Herbig photospheric C and O abundances would remain solar.

\subsection{Inner disk physical conditions and chemistry}
\label{chem}
Another interesting result from this work is the unusual physical conditions within the dust sublimation radius from the Cloudy models. Specifically, the bulk hydrogen and carbon are simultaneously in the neutral atomic phase, contrary to the expectations of a `typical' photon-dominated or x-ray-dominated region, in which carbon photoionizes before H$_2$ photodissociates \citep{maloney1996}. The result found here appears to be a real consequence of the high densities in this region, combined with the lack of solids in the region inside the dust sublimation rim. Specifically, the collisional dissociation rate coefficient for H$_2$-H$_2$ collisions exceeds 10$^{-10}$ cm$^3$s${-1}$ when T>$\sim$8000 K and n$_{H_2}$>10$^{7}$ cm$^{-3}$ \citep{roberge82}. For higher densities, the temperature threshold is lower. Without dust grains, H$_2$ cannot reform efficiently and there is a large neutral H reservoir. 

Then dissociative collisions between CO and H proceed for T>$\sim$ 5000 and n$_H$>10$^{12}$ cm$^{-3}$. Charge transfer at these densities dominates the ionization of H$^+$, producing a fraction of $\sim$10$^{-4}$ of H$^+$ relative to the total hydrogen content. Given the greater abundance of H$^+$ than C$^+$, it is easy for ionized carbon to find an electron to recombine with, keeping the carbon predominantly neutral despite the large $x_e$. Recombination of the small amount of H$^+$ also produces enough UV emission to electronically excite the smaller amount of H$_2$, which then heats the gas to $\sim$7000 K through collisional de-excitation. The low fraction of singly ionized carbon is consistent with the non-detection of emission from its two lowest-lying excited states to its ground state (C$^+$, 1334.532 and 2324.69 \AA) in archival HST COS or STIS spectra of DR Tau. Molecular hydrogen emission has also been found to come from slightly larger disk radii than the CO emission in DR Tau \citep{france2012b}.

The temperatures found from the slab models using CHIANTI are lower than those produced by the Cloudy slab models at the disk midplane. However, the Cloudy models are simple 1D models that only allow heating and cooling in a single dimension. At such close distances to the star, a 2D model becomes necessary to fully explain the propagation of stellar irradiation. If the model is allowed to cool radiatively and vertically, the midplane temperature may be lower than in the 1D versions. Alternatively, the charge exchange rate coefficients, which are ultimately responsible for the heating at the midplane, may have a higher degree of uncertainty at such high densities. I will take this into consideration for the next stage of modeling; the temperature and molecular content at the midplane is particularly important for interpreting the observational signatures of molecular emission coming from this region, which may be weaker than previously anticipated. 

\subsection{Shadowing of the inner dust sublimation rim}
\label{rim}

The Cloudy models indicate that for full disks with midplane densities greater than $n_H\sim$10$^{15}$ cm$^{-3}$, there is a region near the midplane that is optically thick to the stellar radiation. The temperatures found by Cloudy for this hot region are similar to the midplane temperatures found by \citet{muzerolle04} for their optically thick inner disk, although their opacity sources mostly molecular. The optically thick region can have a two-fold effect on observations of the inner disk. First, it will cast a shadow on the inner silicate dust rim. Within this shadow, a thin extension of dust and molecules can potentially exist interior to the dust sublimation rim. A radial extension of midplane dust has the potential to produce an additional contribution to the interferometric signal from the inner rim, with a component originating inside of the minimum possible radius defined by gray-opacity grains at the most refractory dust sublimation temperatures (T$_{sub}$ $\sim$ 1800 K). 

Similarly, if the inner disk gas is optically thick to its own radiation, then it could produce continuum emission, which would also produce a component that appears interior to the sublimation radius. Evidence for a contribution interior to the dust rim can be found in many studies, e.g. \citet{eisner+07,ajay+08, najita+09, lazareff2017}. Understanding whether this component is a radial extension of the inner rim is important for theories of planet formation in the inner disk. \citet{boley+14} appeal to a pressure bump and semi-liquid rocky material at the inner rim in order to build planetary cores rapidly in situ; in their model these cores can eventually become systems of tightly packed inner planets (STIPs) or hot Jupiters without requiring substantial migration \citep{boley2016}.

Further modeling with a 2D model that includes both the gas and a dust rim is needed to determine the spatial extent of an optically thick region, as well as its temperature, to see if it can explain those objects with interferometric signals inside the dust sublimation rim. In principle for my current model with T Tauri stars, the optically thick gas region would be difficult to distinguish from the star, as it would be unresolved and at a similar temperature.

\section{Conclusions}
I have analyzed a set of near-infrared spectra of single, accreting T Tauri stars in the Taurus star-forming region. After determining self-consistently the stellar and accretion parameters using updated distance measurements from GAIA, I extracted the excess from the inner disk. In 18 stars of the 26 star sample, I detect recombination lines of C$^0$ at a $\ge$3$\sigma$ level between 0.9 and 1.5 $\mu$m. For the five stars with a complete set of C$^0$ lines in this region and a high degree of veiling at $\sim$1.25$\mu$m, I modeled the line emission using two slab models: a Cloudy photoionization model to determine the C$^0$ abundance relative to C and an optically thin model with atomic data taken from the CHIANTI database to find the density, temperature, and carbon abundance of the C$^0$ line emitting region. All five stars show evidence for the carbon emission arising in a very dense ($n_H\sim10^{16}$ cm$^{-3}$), warm ($T_e\sim4500$ K), and moderately ionized (log$X_e\sim3.3$) region with a carbon depletion of up to a factor of 42 times with respect to the solar abundance. These parameters match well with predictions of the region interior to the dust sublimation radius, using Cloudy. Analysis of the gas phase carbon depletion at the inner edge of the gas disk implies the following: 

   \begin{enumerate}
      \item Carbon has been depleted from the disk gas onto dust grains and locked into kilometer size bodies somewhere in the disk.
      \item The most likely carrier for the locked carbon is C- and N-rich cometary ices located beyond the N$_2$ snowline, consistent with formation of a proto-Kuiper belt by 1 Myr.
      \item The disk inside the dust sublimation radius should be largely atomic, with a midplane layer that is optically thick to stellar radiation and shadows the dust rim. This, combined with a potentially optically thick emission component from the 4500 K gas, may explain detections of an interferometric component interior to the dust rim in Herbig AeBe stars. 
   \end{enumerate}
   
Expanding this model of the inner disk to take into account the proximity to the star, include a 2D treatment of the stellar radiation, and disentangle the hydrogen emission from the inner disk will be crucial to determining the uncertainties on these abundance measurements and understanding the implications of the physical conditions at the dust rim for forming hot planets in situ. Further higher spectral resolution observations of these C lines and other atomic emission from inside the dust sublimation rim have the potential to be one of the few ways to probe the bulk composition of small planetesimals in the disk. I will explore these possibilities in future work (McClure \& Dominik, in prep).

\begin{acknowledgements}
This publication is part of a project that has received funding from the European Union's Horizon 2020 research and innovation program under the Marie Sklodowska-Curie grant agreement ICED No 749864. It also benefited from suggestions by an anonymous referee. Data was taken as a Visiting Astronomer at the Infrared Telescope Facility, which is operated by the University of Hawaii under contract NNH14CK55B with the National Aeronautics and Space Administration. \end{acknowledgements}

%

\begin{thebibliography}{65}

\bibitem[{{Alexander} {et~al.}(2012){Alexander}, {Bowden}, {Fogel}, {Howard},
  {Herd}, \& {Nittler}}]{alexander2012b}
{Alexander}, C.~M.~O., {Bowden}, R., {Fogel}, M.~L., {et~al.} 2012, Science,
  337, 721

\bibitem[{{Ardila} {et~al.}(2013){Ardila}, {Herczeg}, {Gregory}, {Ingleby},
  {France}, {Brown}, {Edwards}, {Johns-Krull}, {Linsky}, {Yang}, {Valenti},
  {Abgrall}, {Alexander}, {Bergin}, {Bethell}, {Brown}, {Calvet}, {Espaillat},
  {Hillenbrand}, {Hussain}, {Roueff}, {Schindhelm}, \& {Walter}}]{ardila2013}
{Ardila}, D.~R., {Herczeg}, G.~J., {Gregory}, S.~G., {et~al.} 2013, The
  Astrophysical Journal Supplement Series, 207, 1

\bibitem[{{Asplund} {et~al.}(2009){Asplund}, {Grevesse}, {Sauval}, \&
  {Scott}}]{asplund2009}
{Asplund}, M., {Grevesse}, N., {Sauval}, A.~J., \& {Scott}, P. 2009, \araa, 47,
  481

\bibitem[{{Bailer-Jones} {et~al.}(2018){Bailer-Jones}, {Rybizki}, {Fouesneau},
  {Mantelet}, \& {Andrae}}]{bailer-jones18}
{Bailer-Jones}, C.~A.~L., {Rybizki}, J., {Fouesneau}, M., {Mantelet}, G., \&
  {Andrae}, R. 2018, ArXiv e-prints, arXiv:1804.10121

\bibitem[{{Banzatti} \& {Pontoppidan}(2015)}]{banzatti2015}
{Banzatti}, A. \& {Pontoppidan}, K.~M. 2015, \apj, 809, 167

\bibitem[{{Bergin} {et~al.}(2013){Bergin}, {Cleeves}, {Gorti}, {Zhang},
  {Blake}, {Green}, {Andrews}, {Evans}, {Henning}, {{\"O}berg}, {Pontoppidan},
  {Qi}, {Salyk}, \& {van Dishoeck}}]{bergin+13}
{Bergin}, E.~A., {Cleeves}, L.~I., {Gorti}, U., {et~al.} 2013, \nat, 493, 644

\bibitem[{{Birnstiel} {et~al.}(2012){Birnstiel}, {Klahr}, \&
  {Ercolano}}]{birnstiel+12}
{Birnstiel}, T., {Klahr}, H., \& {Ercolano}, B. 2012, \aap, 539, A148

\bibitem[{{Boley} {et~al.}(2016){Boley}, {Granados Contreras}, \&
  {Gladman}}]{boley2016}
{Boley}, A.~C., {Granados Contreras}, A.~P., \& {Gladman}, B. 2016, \apj, 817,
  L17

\bibitem[{{Boley} {et~al.}(2014){Boley}, {Morris}, \& {Ford}}]{boley+14}
{Boley}, A.~C., {Morris}, M.~A., \& {Ford}, E.~B. 2014, \apjl, 792, L27

\bibitem[{{Brown} {et~al.}(2013){Brown}, {Pontoppidan}, {van Dishoeck},
  {Herczeg}, {Blake}, \& {Smette}}]{brown2013}
{Brown}, J.~M., {Pontoppidan}, K.~M., {van Dishoeck}, E.~F., {et~al.} 2013,
  \apj, 770, 94

\bibitem[{{Castelli} \& {Kurucz}(2004)}]{castelli2004}
{Castelli}, F. \& {Kurucz}, R.~L. 2004, ArXiv Astrophysics e-prints
  [\eprint{astro-ph/0405087}]

\bibitem[{{Cushing} {et~al.}(2004){Cushing}, {Vacca}, \& {Rayner}}]{cushing+04}
{Cushing}, M.~C., {Vacca}, W.~D., \& {Rayner}, J.~T. 2004, \pasp, 116, 362

\bibitem[{{D'Alessio} {et~al.}(2006){D'Alessio}, {Calvet}, {Hartmann},
  {Franco-Hern{\'a}ndez}, \& {Serv{\'{\i}}n}}]{dalessio+06}
{D'Alessio}, P., {Calvet}, N., {Hartmann}, L., {Franco-Hern{\'a}ndez}, R., \&
  {Serv{\'{\i}}n}, H. 2006, \apj, 638, 314

\bibitem[{{Dartois} {et~al.}(2017){Dartois}, {Chabot}, {Pino}, {B{\'e}roff},
  {Godard}, {Severin}, {Bender}, \& {Trautmann}}]{dartois2017}
{Dartois}, E., {Chabot}, M., {Pino}, T., {et~al.} 2017, \aap, 599, A130

\bibitem[{{Dere} {et~al.}(1997){Dere}, {Landi}, {Mason}, {Monsignori Fossi}, \&
  {Young}}]{dere+97}
{Dere}, K.~P., {Landi}, E., {Mason}, H.~E., {Monsignori Fossi}, B.~C., \&
  {Young}, P.~R. 1997, \aaps, 125, 149

\bibitem[{{Eisner} {et~al.}(2014){Eisner}, {Hillenbrand}, \&
  {Stone}}]{eisner2014}
{Eisner}, J.~A., {Hillenbrand}, L.~A., \& {Stone}, J.~M. 2014, \mnras, 443,
  1916

\bibitem[{{Eisner} {et~al.}(2007){Eisner}, {Hillenbrand}, {White}, {Bloom},
  {Akeson}, \& {Blake}}]{eisner+07}
{Eisner}, J.~A., {Hillenbrand}, L.~A., {White}, R.~J., {et~al.} 2007, \apj,
  669, 1072

\bibitem[{{Engrand} {et~al.}(2018){Engrand}, {Charon}, {Duprat}, {Dartois},
  {Leroux}, {Benzerara}, {Le Guillou}, {Bernard}, {Swaraj}, {Belkhou},
  {Delauche}, {Godard}, \& {Aug{\'e}}}]{engrand2018a}
{Engrand}, C., {Charon}, E., {Duprat}, J., {et~al.} 2018, in Lunar and
  Planetary Science Conference, Vol.~49, Lunar and Planetary Science
  Conference, 2015

\bibitem[{{Escalante} \& {Victor}(1990)}]{escalante90}
{Escalante}, V. \& {Victor}, G.~A. 1990, \apjs, 73, 513

\bibitem[{{Espaillat} {et~al.}(2014){Espaillat}, {Muzerolle}, {Najita},
  {Andrews}, {Zhu}, {Calvet}, {Kraus}, {Hashimoto}, {Kraus}, \&
  {D'Alessio}}]{espaillat+14}
{Espaillat}, C., {Muzerolle}, J., {Najita}, J., {et~al.} 2014, Protostars and
  Planets VI, 497

\bibitem[{{Favre} {et~al.}(2013){Favre}, {Cleeves}, {Bergin}, {Qi}, \&
  {Blake}}]{favre+13}
{Favre}, C., {Cleeves}, L.~I., {Bergin}, E.~A., {Qi}, C., \& {Blake}, G.~A.
  2013, \apjl, 776, L38

\bibitem[{{Ferland} {et~al.}(2017){Ferland}, {Chatzikos}, {Guzm{\'a}n},
  {Lykins}, {van Hoof}, {Williams}, {Abel}, {Badnell}, {Keenan}, {Porter}, \&
  {Stancil}}]{ferland2017}
{Ferland}, G.~J., {Chatzikos}, M., {Guzm{\'a}n}, F., {et~al.} 2017, \rmxaa, 53,
  385

\bibitem[{{France} {et~al.}(2012){France}, {Schindhelm}, {Herczeg}, {Brown},
  {Abgrall}, {Alexander}, {Bergin}, {Brown}, {Linsky}, {Roueff}, \&
  {Yang}}]{france2012b}
{France}, K., {Schindhelm}, E., {Herczeg}, G.~J., {et~al.} 2012, \apj, 756, 171

\bibitem[{{Fray} {et~al.}(2017){Fray}, {Bardyn}, {Cottin}, {Baklouti},
  {Briois}, {Engrand}, {Fischer}, {Hornung}, {Isnard}, {Langevin}, {Lehto}, {Le
  Roy}, {Mellado}, {Merouane}, {Modica}, {Orthous-Daunay}, {Paquette},
  {Ryn{\"o}}, {Schulz}, {Sil{\'e}n}, {Siljestr{\"o}m}, {Stenzel}, {Thirkell},
  {Varmuza}, {Zaprudin}, {Kissel}, \& {Hilchenbach}}]{fray2017}
{Fray}, N., {Bardyn}, A., {Cottin}, H., {et~al.} 2017, \mnras, 469, S506

\bibitem[{{Gammie}(1996)}]{gammie1996}
{Gammie}, C.~F. 1996, \apj, 457, 355

\bibitem[{Haris \& Kramida(2017)}]{haris_2017}
Haris, K. \& Kramida, A. 2017, The Astrophysical Journal Supplement Series,
  233, 16

\bibitem[{{Ingleby} {et~al.}(2013){Ingleby}, {Calvet}, {Herczeg}, {Blaty},
  {Walter}, {Ardila}, {Alexander}, {Edwards}, {Espaillat}, {Gregory},
  {Hillenbrand}, \& {Brown}}]{ingleby+13}
{Ingleby}, L., {Calvet}, N., {Herczeg}, G., {et~al.} 2013, \apj, 767, 112

\bibitem[{{Kama} {et~al.}(2016){Kama}, {Bruderer}, {van Dishoeck},
  {Hogerheijde}, {Folsom}, {Miotello}, {Fedele}, {Belloche}, {G{\"u}sten}, \&
  {Wyrowski}}]{kama+16b}
{Kama}, M., {Bruderer}, S., {van Dishoeck}, E.~F., {et~al.} 2016, \aap, 592,
  A83

\bibitem[{{Kargel} \& {Lewis}(1993)}]{kargel1993}
{Kargel}, J.~S. \& {Lewis}, J.~S. 1993, \icarus, 105, 1

\bibitem[{{Kenyon} {et~al.}(1994){Kenyon}, {Dobrzycka}, \&
  {Hartmann}}]{kenyon+94}
{Kenyon}, S.~J., {Dobrzycka}, D., \& {Hartmann}, L. 1994, \aj, 108, 1872

\bibitem[{{Kenyon} \& {Hartmann}(1995)}]{kh95}
{Kenyon}, S.~J. \& {Hartmann}, L. 1995, \apjs, 101, 117

\bibitem[{{Kingdon} \& {Ferland}(1996)}]{kingdon1996}
{Kingdon}, J.~B. \& {Ferland}, G.~J. 1996, \apjs, 106, 205

\bibitem[{{Kingdon} \& {Ferland}(1999)}]{kingdon1999}
{Kingdon}, J.~B. \& {Ferland}, G.~J. 1999, \apj, 516, L107

\bibitem[{Kramida {et~al.}(2018)Kramida, {Yu.~Ralchenko}, Reader, \& {and NIST
  ASD Team}}]{NIST_ASD}
Kramida, A., {Yu.~Ralchenko}, Reader, J., \& {and NIST ASD Team}. 2018, {NIST
  Atomic Spectra Database (ver. 5.5.2), [Online]. Available:
  {\tt{https://physics.nist.gov/asd}} [2018, January 25]. National Institute of
  Standards and Technology, Gaithersburg, MD.}

\bibitem[{{Landi} {et~al.}(2013){Landi}, {Young}, {Dere}, {Del Zanna}, \&
  {Mason}}]{landi+13}
{Landi}, E., {Young}, P.~R., {Dere}, K.~P., {Del Zanna}, G., \& {Mason}, H.~E.
  2013, \apj, 763, 86

\bibitem[{{Lazareff} {et~al.}(2017){Lazareff}, {Berger}, {Kluska}, {Le
  Bouquin}, {Benisty}, {Malbet}, {Koen}, {Pinte}, {Thi}, {Absil}, {Baron},
  {Delboulb{\'e}}, {Duvert}, {Isella}, {Jocou}, {Juhasz}, {Kraus}, {Lachaume},
  {M{\'e}nard}, {Millan-Gabet}, {Monnier}, {Moulin}, {Perraut}, {Rochat},
  {Soulez}, {Tallon}, {Thi{\'e}baut}, {Traub}, \& {Zins}}]{lazareff2017}
{Lazareff}, B., {Berger}, J.~P., {Kluska}, J., {et~al.} 2017, \aap, 599, A85

\bibitem[{{Long} {et~al.}(2018){Long}, {Pinilla}, {Herczeg}, {Harsono},
  {Dipierro}, {Pascucci}, {Hendler}, {Tazzari}, {Ragusa}, {Salyk}, {Edwards},
  {Lodato}, {van de Plas}, {Johnstone}, {Liu}, {Boehler}, {Cabrit}, {Manara},
  {Menard}, {Mulders}, {Nisini}, {Fischer}, {Rigliaco}, {Banzatti}, {Avenhaus},
  \& {Gully-Santiago}}]{long2018b}
{Long}, F., {Pinilla}, P., {Herczeg}, G.~J., {et~al.} 2018, \apj, 869, 17

\bibitem[{{Madhusudhan}(2012)}]{madhusudhan12b}
{Madhusudhan}, N. 2012, \apj, 758, 36

\bibitem[{{Maloney} {et~al.}(1996){Maloney}, {Hollenbach}, \&
  {Tielens}}]{maloney1996}
{Maloney}, P.~R., {Hollenbach}, D.~J., \& {Tielens}, A.~G.~G.~M. 1996, \apj,
  466, 561

\bibitem[{{McClure} {et~al.}(2016){McClure}, {Bergin}, {Cleeves}, {van
  Dishoeck}, {Blake}, {Evans}, {Green}, {Henning}, {{\"O}berg}, {Pontoppidan},
  \& {Salyk}}]{mcclure+16}
{McClure}, M.~K., {Bergin}, E.~A., {Cleeves}, L.~I., {et~al.} 2016, \apj, 831,
  167

\bibitem[{{McClure} {et~al.}(2013{\natexlab{a}}){McClure}, {Calvet},
  {Espaillat}, {Hartmann}, {Hern{\'a}ndez}, {Ingleby}, {Luhman}, {D'Alessio},
  \& {Sargent}}]{mcclure+13a}
{McClure}, M.~K., {Calvet}, N., {Espaillat}, C., {et~al.} 2013{\natexlab{a}},
  \apj, 769, 73

\bibitem[{{McClure} {et~al.}(2013{\natexlab{b}}){McClure}, {D'Alessio},
  {Calvet}, {Espaillat}, {Hartmann}, {Sargent}, {Watson}, {Ingleby}, \&
  {Hern{\'a}ndez}}]{mcclure+13b}
{McClure}, M.~K., {D'Alessio}, P., {Calvet}, N., {et~al.} 2013{\natexlab{b}},
  \apj, 775, 114

\bibitem[{{McClure} {et~al.}(2015){McClure}, {Espaillat}, {Calvet}, {Bergin},
  {D'Alessio}, {Watson}, {Manoj}, {Sargent}, \& {Cleeves}}]{mcclure+15}
{McClure}, M.~K., {Espaillat}, C., {Calvet}, N., {et~al.} 2015, \apj, 799, 162

\bibitem[{{Miotello} {et~al.}(2017){Miotello}, {van Dishoeck}, {Williams},
  {Ansdell}, {Guidi}, {Hogerheijde}, {Manara}, {Tazzari}, {Testi}, {van der
  Marel}, \& {van Terwisga}}]{miotello2017}
{Miotello}, A., {van Dishoeck}, E.~F., {Williams}, J.~P., {et~al.} 2017, \aap,
  599, A113

\bibitem[{{Musiolik} {et~al.}(2016){Musiolik}, {Teiser}, {Jankowski}, \&
  {Wurm}}]{musiolik2016}
{Musiolik}, G., {Teiser}, J., {Jankowski}, T., \& {Wurm}, G. 2016, \apj, 818,
  16

\bibitem[{{Muzerolle} {et~al.}(2004){Muzerolle}, {D'Alessio}, {Calvet}, \&
  {Hartmann}}]{muzerolle04}
{Muzerolle}, J., {D'Alessio}, P., {Calvet}, N., \& {Hartmann}, L. 2004, \apj,
  617, 406

\bibitem[{{Muzerolle} {et~al.}(2003){Muzerolle}, {Hillenbrand}, {Calvet},
  {Brice{\~n}o}, \& {Hartmann}}]{muzerolle03a}
{Muzerolle}, J., {Hillenbrand}, L., {Calvet}, N., {Brice{\~n}o}, C., \&
  {Hartmann}, L. 2003, \apj, 592, 266

\bibitem[{{Najita} \& {Bergin}(2018)}]{najita_bergin2018}
{Najita}, J.~R. \& {Bergin}, E.~A. 2018, \apj, 864, 168

\bibitem[{{Najita} {et~al.}(2009){Najita}, {Doppmann}, {Carr}, {Graham}, \&
  {Eisner}}]{najita+09}
{Najita}, J.~R., {Doppmann}, G.~W., {Carr}, J.~S., {Graham}, J.~R., \&
  {Eisner}, J.~A. 2009, \apj, 691, 738

\bibitem[{{Piso} {et~al.}(2016){Piso}, {Pegues}, \& {{\"O}berg}}]{piso2016}
{Piso}, A.-M.~A., {Pegues}, J., \& {{\"O}berg}, K.~I. 2016, \apj, 833, 203

\bibitem[{{Pontoppidan} {et~al.}(2014){Pontoppidan}, {Salyk}, {Bergin},
  {Brittain}, {Marty}, {Mousis}, \& {{\"O}berg}}]{pontoppidan+14}
{Pontoppidan}, K.~M., {Salyk}, C., {Bergin}, E.~A., {et~al.} 2014, Protostars
  and Planets VI, 363

\bibitem[{{Rab} {et~al.}(2018){Rab}, {G{\"u}del}, {Woitke}, {Kamp}, {Thi},
  {Min}, {Aresu}, \& {Meijerink}}]{rab2017}
{Rab}, C., {G{\"u}del}, M., {Woitke}, P., {et~al.} 2018, \aap, 609, A91

\bibitem[{{Rayner} {et~al.}(2009){Rayner}, {Cushing}, \& {Vacca}}]{rayner+09}
{Rayner}, J.~T., {Cushing}, M.~C., \& {Vacca}, W.~D. 2009, \apjs, 185, 289

\bibitem[{{Rayner} {et~al.}(2003){Rayner}, {Toomey}, {Onaka}, {Denault},
  {Stahlberger}, {Vacca}, {Cushing}, \& {Wang}}]{rayner+03}
{Rayner}, J.~T., {Toomey}, D.~W., {Onaka}, P.~M., {et~al.} 2003, \pasp, 115,
  362

\bibitem[{{Roberge} \& {Dalgarno}(1982)}]{roberge82}
{Roberge}, W. \& {Dalgarno}, A. 1982, \apj, 255, 176

\bibitem[{{Salyk} {et~al.}(2011){Salyk}, {Blake}, {Boogert}, \&
  {Brown}}]{salyk2011}
{Salyk}, C., {Blake}, G.~A., {Boogert}, A.~C.~A., \& {Brown}, J.~M. 2011, \apj,
  743, 112

\bibitem[{{Schwarz} {et~al.}(2016){Schwarz}, {Bergin}, {Cleeves}, {Blake},
  {Zhang}, {{\"O}berg}, {van Dishoeck}, \& {Qi}}]{schwarz+16a}
{Schwarz}, K.~R., {Bergin}, E.~A., {Cleeves}, L.~I., {et~al.} 2016, \apj, 823,
  91

\bibitem[{{Shaw} {et~al.}(2005){Shaw}, {Ferland}, {Abel}, {Stancil}, \& {van
  Hoof}}]{shaw2005}
{Shaw}, G., {Ferland}, G.~J., {Abel}, N.~P., {Stancil}, P.~C., \& {van Hoof},
  P.~A.~M. 2005, \apj, 624, 794

\bibitem[{{Siess} {et~al.}(2000){Siess}, {Dufour}, \& {Forestini}}]{siess+00}
{Siess}, L., {Dufour}, E., \& {Forestini}, M. 2000, \aap, 358, 593

\bibitem[{{Tannirkulam} {et~al.}(2008){Tannirkulam}, {Monnier}, {Millan-Gabet},
  {Harries}, {Pedretti}, {ten Brummelaar}, {McAlister}, {Turner}, {Sturmann},
  \& {Sturmann}}]{ajay+08}
{Tannirkulam}, A., {Monnier}, J.~D., {Millan-Gabet}, R., {et~al.} 2008, \apjl,
  677, L51

\bibitem[{{Vacca} {et~al.}(2003){Vacca}, {Cushing}, \& {Rayner}}]{vacca+03}
{Vacca}, W.~D., {Cushing}, M.~C., \& {Rayner}, J.~T. 2003, \pasp, 115, 389

\bibitem[{{Vacca} \& {Sandell}(2011)}]{vacca_sandell11}
{Vacca}, W.~D. \& {Sandell}, G. 2011, \apj, 732, 8

\bibitem[{{van 't Hoff} {et~al.}(2017){van 't Hoff}, {Walsh}, {Kama},
  {Facchini}, \& {van Dishoeck}}]{vanthoff2017}
{van 't Hoff}, M.~L.~R., {Walsh}, C., {Kama}, M., {Facchini}, S., \& {van
  Dishoeck}, E.~F. 2017, \aap, 599, A101

\bibitem[{{Yu} {et~al.}(2017){Yu}, {Evans}, {Dodson-Robinson}, {Willacy}, \&
  {Turner}}]{yu2017}
{Yu}, M., {Evans}, Neal~J., I., {Dodson-Robinson}, S.~E., {Willacy}, K., \&
  {Turner}, N.~J. 2017, \apj, 841, 39

\bibitem[{{Zhang} {et~al.}(2017){Zhang}, {Bergin}, {Blake}, {Cleeves}, \&
  {Schwarz}}]{zhang2017}
{Zhang}, K., {Bergin}, E.~A., {Blake}, G.~A., {Cleeves}, L.~I., \& {Schwarz},
  K.~R. 2017, Nature Astronomy, 1, 0130

\end{thebibliography}


\begin{appendix} 
\onecolumn
\section{Atomic line data} 
\label{atomicdata}
Most of the C$^0$ line complexes are composed of multiple, blended lines.  For each individual line in these complexes, Table \ref{nistinfo} lists the wavelengths in vacuum, Einstein A coefficients, level energies, and lower and upper configuration terms distilled from the NIST Atomic Spectra Database \citep{NIST_ASD}. The line complexes given around 0.96 $\mu$m and 1.26 $\mu$m are very weak and only visually suggested by the spectra upon comparison between targets. Their fluxes are not measured or used for any of the analysis here.

\begin{table*}
\caption{Complete observed neutral carbon line list}\label{starbursts}
\centering
\begin{tabular}{cccccc}
\hline\hline                 
wavelength  & $A_{upper, lower}$ & $E_{lower}$ & $E_{upper}$ & Lower level & Upper level \\    
($\mu$m)    &     (s$^{-1}$) & (eV)  & (eV) & configuration, term, J & configuration, term, J \\
\hline                        
0.906392	&	7.31$\times$10$^{6}$	&	7.482772	&	8.850659	&	2s22p3s, 3P$^\circ$, 1	&	2s22p3p, 3P, 2	\\
0.906496	&	9.48$\times$10$^{6}$	&	7.480392	&	8.84812	&	2s22p3s, 3P$^\circ$, 0	&	2s22p3p, 3P, 1	\\
0.908077	&	7.07$\times$10$^{6}$	&	7.482772	&	8.84812	&	2s22p3s, 3P$^\circ$, 1	&	2s22p3p, 3P, 1	\\
0.9091	&	3.00$\times$10$^{7}$	&	7.482772	&	8.846584	&	2s22p3s, 3P$^\circ$, 1	&	2s22p3p, 3P, 0	\\
0.909733	&	2.28$\times$10$^{7}$	&	7.487795	&	8.850659	&	2s22p3s, 3P$^\circ$, 2	&	2s22p3p, 3P, 2	\\
0.91143	&	1.35$\times$10$^{7}$	&	7.487795	&	8.84812	&	2s22p3s, 3P$^\circ$, 2	&	2s22p3p, 3P, 1	\\
\hline																			
0.940831	&	2.91$\times$10$^{7}$	&	7.684766	&	9.002582	&	2s22p3s, 1P$^\circ$, 1	&	2s22p3p, 1D, 2	\\
\hline																			
0.96056648	&	3.10$\times$10$^{6}$	&	7.48039408 & 8.77113462 &	2s22p3s, 3P$^\circ$, 0	&	2s22p3p, 3S, 1	\\
0.96234214	&	8.60$\times$10$^{6}$	&	7.48277591 & 8.77113462 &	2s22p3s, 3P$^\circ$, 1	&	2s22p3p, 3S, 1	\\
0.96610867	&	1.25$\times$10$^{7}$	&	7.48779862 & 8.77113462 &	2s22p3s, 3P$^\circ$, 2	&	2s22p3p, 3S, 1	\\
\hline																			
1.068601	&	1.40$\times$10$^{7}$	&	7.482772	&	8.64302	&	2s22p3s, 3P$^\circ$, 1	&	2s22p3p, 3D, 2	\\
1.068827	&	1.04$\times$10$^{7}$	&	7.480392	&	8.640394	&	2s22p3s, 3P$^\circ$, 0	&	2s22p3p, 3D, 1	\\
1.069418	&	1.84$\times$10$^{7}$	&	7.487795	&	8.647158	&	2s22p3s, 3P$^\circ$, 2	&	2s22p3p, 3D, 3	\\
1.071025	&	7.53$\times$10$^{6}$	&	7.482772	&	8.640394	&	2s22p3s, 3P$^\circ$, 1	&	2s22p3p, 3D, 1	\\
1.073247	&	4.40$\times$10$^{6}$	&	7.487795	&	8.64302	&	2s22p3s, 3P$^\circ$, 2	&	2s22p3p, 3D, 2	\\
\hline																			
1.162247	&	4.39$\times$10$^{6}$	&	8.640394	&	9.707156	&	2s22p3p, 3D, 1	&	2s22p3d, 3D$^\circ$, 1	\\
1.163201	&	5.42$\times$10$^{6}$	&	8.64302	&	9.708909	&	2s22p3p, 3D, 2	&	2s22p3d, 3D$^\circ$, 2	\\
1.166287	&	7.48$\times$10$^{6}$	&	8.647158	&	9.710225	&	2s22p3p, 3D, 3	&	2s22p3d, 3D$^\circ$, 3	\\
1.167734	&	1.57$\times$10$^{6}$	&	8.647158	&	9.708909	&	2s22p3p, 3D, 3	&	2s22p3d, 3D$^\circ$, 2	\\
\hline																			
1.175144	&	2.29$\times$10$^{7}$	&	8.640394	&	9.695449	&	2s22p3p, 3D, 1	&	2s22p3d, 3F$^\circ$, 2	\\
1.175654	&	2.63$\times$10$^{7}$	&	8.647158	&	9.701756	&	2s22p3p, 3D, 3	&	2s22p3d, 3F$^\circ$, 4	\\
1.175798	&	2.40$\times$10$^{7}$	&	8.64302	&	9.697487	&	2s22p3p, 3D, 2	&	2s22p3d, 3F$^\circ$, 3	\\
\hline																			
1.189617	&	8.31$\times$10$^{6}$	&	8.64302	&	9.685241	&	2s22p3p, 3D, 2	&	2s22p4s, 3P$^\circ$, 1	\\
1.189901	&	9.24$\times$10$^{6}$	&	8.647158	&	9.689128	&	2s22p3p, 3D, 3	&	2s22p4s, 3P$^\circ$, 2	\\
\hline																			
1.255292	&	3.84$\times$10$^{6}$	&	8.846584	&	9.834275	&	2s22p3p, 3P, 0	&	2s22p3d, 3P$^\circ$, 1	\\
1.256556	&	1.27$\times$10$^{7}$	&	8.84812	&	9.834818	&	2s22p3p, 3P, 1	&	2s22p3d, 3P$^\circ$, 0	\\
1.257248	&	3.55$\times$10$^{6}$	&	8.84812	&	9.834275	&	2s22p3p, 3P, 1	&	2s22p3d, 3P$^\circ$, 1	\\
1.258503	&	2.45$\times$10$^{6}$	&	8.84812	&	9.833292	&	2s22p3p, 3P, 1	&	2s22p3d, 3P$^\circ$, 2	\\
1.260494	&	5.05$\times$10$^{6}$	&	8.850659	&	9.834275	&	2s22p3p, 3P, 2	&	2s22p3d, 3P$^\circ$, 1	\\
1.261755	&	9.39$\times$10$^{6}$	&	8.850659	&	9.833292	&	2s22p3p, 3P, 2	&	2s22p3d, 3P$^\circ$, 2	\\
\hline																			
1.454649	&	8.55$\times$10$^{6}$	&	7.684766	&	8.537097	&	2s22p3s, 1P$^\circ$, 1	&	2s22p3p, 1P, 1	\\
\hline
\hline
\label{nistinfo}
\end{tabular}
\\
\textbf{Notes:} Horizontal lines separate the different line complexes seen in the spectra. \\
NIST ASD can be accessed here: https://physics.nist.gov/PhysRefData/ASD/lines$\_$form.html?
\end{table*}




\section{Outer radii assumed for slab models}
\label{outradapp}

The radial 1D Cloudy model can only cool in one dimension, so it never reaches temperatures where CO would dominate over C$^0$. Therefore, it is necessary to infer the outer radius of the C$^0$ emission from literature measurements of H$^0$, CO, and the dust sublimation rim. 

Two of the disks, DR Tau and DO Tau, have all of these measurements. In H$^0$, the outermost radius contributing to Br$\gamma$ emission has been measured interferometrically, at 0.05 and 0.1 AU \citep{eisner2014}, respectively.  A radius of the peak emission from a hot, broad component of CO has been measured kinematically from CRIRES spectra of the CO fundamental rovibrational band at $\sim$4.7 $\mu$m \citep{banzatti2015}, at 0.06 and 0.12 AU, respectively. This peak emission comes from just outside of the Br$\gamma$ outer radius. In the case of DO Tau, weaker CO emission may extend to an innermost CO radius of 0.03 AU \citep{salyk2011}. For both DR Tau and DO Tau, the dust sublimation rim has been interferometrically measured at 0.12 and 0.18 AU, respectively, outside of the peak CO emission radius \citep{eisner2014}. For DO Tau, this rim radius is also consistent with SED modeling of the inner rim excess \citep{mcclure+15}. For these two disks, I assume the H$^0$ Br$\gamma$ outer radius as the outer radius for C$^0$. Realistically, a large population of H$^0$ is necessary in order to produce free electrons from the formation of H$^+$, which can then recombine with C$^+$ to maintain the neutral carbon population and produce these lines.

The other three disks in the sample lack the complete set of radii observations that DR Tau and DO Tau have. Since the carbon abundance required to fit the observed flux will increase if the emitting region is made smaller, I have chosen to use the smallest radii associated with CW Tau, FZ Tau, and DL Tau in order to present a conservative upper limit on the carbon abundance.

\begin{table}
\caption{Slab radii}             
\label{table_rad}      
\centering                          
\begin{tabular}{c c c c c c c}        
\hline\hline                 
\\
Star  & R$_{slab}$ & R$_{H Br\gamma, out}$ & R$_{CO, inner}$ & R$_{CO, BC peak}$ & R$_{dust, inter}$ & R$_{dust, SED}$ \\    
    &     [AU] & [AU] & [AU]  & [AU] & [AU] & [AU] \\
\hline                        
CW Tau & 0.04 & -                                &  0.04$^{a, 1}$     &  0.32$^b$ &  -                                   &  - \\
DR Tau  & 0.05 & 0.05$\pm$0.01 $^c$ &  0.2-0.5$^d$      &  0.06$^b$ &   0.12$\pm$0.1 $^c$    &   -  \\
DO Tau  & 0.1   & 0.10$\pm$0.01 $^c$ &  0.03-0.05$^d$  &  0.12$^b$ &   0.18$\pm$0.01 $^c$  &   0.17$^e$ \\ 
FZ Tau   & 0.1   & -                                &  0.1$^{a, 2}$        &  1.0$^b$  &    -                                 &   0.26$^f$ \\
DL Tau  &  0.05 & -                                &  0.01-0.05$^d$  &  -             &  -                                   &  -  \\

\label{router}
\end{tabular}
\\
\textbf{References:} (a) \citet{brown2013}, (b) \citet{banzatti2015}, (c) \citet{eisner2014}, (d) \citet{salyk2011}, (e) \citet{mcclure+15}, (f) \citet{mcclure+16} \\
\textbf{Notes:} (1) From $R$sin$i$, assuming an inclination of 28 degrees from \citet{banzatti2015}. (2) From $R$sin$i$, assuming an inclination of 38 degrees from \citet{banzatti2015}. 
\end{table}



\section{Line optical depth verification}
\label{optdepthapp}

The optical depth of the C$^0$ lines can be expressed as:

\begin{equation}
\tau_{u,l}=\frac{n_u}{n_c}X_C N_H A_{ul} \frac{\lambda^3}{8\pi\Delta v}(e^{\frac{h\nu}{kT}}-1)
\end{equation}

\noindent where $\Delta v$ is the width of the line in km s$^{-1}$, $A_{ul}$ is the Einstein transition probability in s$^{-1}$, and N$_u$, the column density of C$^0$ in the upper level of the transition, is the product of the first three terms: the fraction of C atoms in the upper level state at a given $n_e$ and $T_e$, the carbon abundance relative to hydrogen, and the hydrogen column density. I calculate $\tau_{u,l}$ using the $n_e$, $T_e$, $n_H$, and $X_C$ parameters found for each disk in the main paper, with $n_u/n_c$ from calculations using the CHIANTI atomic line database, and assuming a gaussian density profile to estimate $N_H$.  Note that all of the lines are optically thin for each star except for CW Tau; for this star, the density is high enough that the main line in the 0.91 $\mu$m complex and the two brightest lines in the 1.07 $\mu$m complex have $\tau_{line} \sim$1.

\begin{table*}[!b]
\caption{Line optical depths}             
\label{table_opt}      
\centering                          
\begin{tabular}{c c c c c c}        
\hline\hline                 
\\
$\tau$  & CW Tau & DR Tau & FZ Tau & DO Tau & DL Tau  \\    
\hline \\
0.9064 & 3.2$\times$10$^{-1}$ & 7.7$\times$10$^{-2}$ & 3.1$\times$10$^{-3}$ & 3.6$\times$10$^{-2}$ & 5.3$\times$10$^{-2}$ \\
0.9065 & 2.5$\times$10$^{-1}$ & 5.9$\times$10$^{-2}$ & 2.3$\times$10$^{-3}$ & 2.8$\times$10$^{-2}$ & 4.1$\times$10$^{-2}$ \\
0.9081 & 1.9$\times$10$^{-1}$ & 4.4$\times$10$^{-2}$ & 1.7$\times$10$^{-3}$ & 2.1$\times$10$^{-2}$ & 3.0$\times$10$^{-2}$ \\
0.9091 & 2.7$\times$10$^{-1}$ & 6.3$\times$10$^{-2}$ & 2.5$\times$10$^{-3}$ & 3.0$\times$10$^{-2}$ & 4.4$\times$10$^{-2}$ \\
0.9097 & 1.0 & 2.4$\times$10$^{-1}$ & 9.5$\times$10$^{-3}$ & 1.1$\times$10$^{-1}$ & 1.7$\times$10$^{-1}$ \\
0.9114 & 3.5$\times$10$^{-1}$ & 8.4$\times$10$^{-2}$ & 3.3$\times$10$^{-3}$ & 4.0$\times$10$^{-2}$ & 5.8$\times$10$^{-2}$ \\
0.9408 & 8.4$\times$10$^{-1}$ & 2.0$\times$10$^{-1}$ & 7.7$\times$10$^{-3}$ & 1.0$\times$10$^{-1}$ & 1.5$\times$10$^{-1}$ \\
1.0686 & 1.3 & 3.2$\times$10$^{-1}$ & 1.3$\times$10$^{-2}$ & 1.5$\times$10$^{-1}$ & 2.3$\times$10$^{-1}$ \\
1.0688 & 7.1$\times$10$^{-1}$ & 1.7$\times$10$^{-1}$ & 7.2$\times$10$^{-3}$ & 8.4$\times$10$^{-2}$ & 1.2$\times$10$^{-1}$ \\
1.0694 & 1.5 & 4.8$\times$10$^{-1}$ & 2.8$\times$10$^{-2}$ & 3.2$\times$10$^{-1}$ & 5.1$\times$10$^{-1}$ \\
1.0710 & 5.2$\times$10$^{-1}$ & 1.3$\times$10$^{-1}$ & 5.2$\times$10$^{-3}$ & 6.1$\times$10$^{-2}$ & 9.0$\times$10$^{-2}$ \\
1.0732 & 4.1$\times$10$^{-1}$ & 1.0$\times$10$^{-1}$ & 4.1$\times$10$^{-3}$ & 4.9$\times$10$^{-2}$ & 7.2$\times$10$^{-2}$ \\
1.1751 & 4.4$\times$10$^{-2}$ & 5.9$\times$10$^{-3}$ & 1.1$\times$10$^{-4}$ & 2.2$\times$10$^{-3}$ & 2.7$\times$10$^{-3}$ \\
1.1757 & 1.0$\times$10$^{-1}$ & 1.5$\times$10$^{-2}$ & 3.2$\times$10$^{-4}$ & 6.4$\times$10$^{-3}$ & 8.3$\times$10$^{-3}$ \\
1.1758 & 5.6$\times$10$^{-2}$ & 6.9$\times$10$^{-3}$ & 1.2$\times$10$^{-4}$ & 2.3$\times$10$^{-3}$ & 2.9$\times$10$^{-3}$ \\
1.4546 & 3.9$\times$10$^{-1}$ & 9.4$\times$10$^{-2}$ & 3.8$\times$10$^{-3}$ & 5.2$\times$10$^{-2}$ & 7.7$\times$10$^{-2}$ \\
\hline \\                       
\label{opttab}
\end{tabular}
\\
\end{table*}

\end{appendix}




\end{document}